\documentclass{article}

\usepackage{arxiv}

\usepackage[utf8]{inputenc} 
\usepackage[T1]{fontenc}    
\usepackage{hyperref}       
\usepackage{url}            
\usepackage{booktabs}       
\usepackage{amsfonts}       
\usepackage{nicefrac}       
\usepackage{microtype}      
\usepackage{lipsum}
\usepackage[table]{xcolor}
\usepackage{graphicx}
\usepackage{comment}
\usepackage{float}
\usepackage{natbib} 

\newcommand{\argmax}[1]{\underset{#1}{\operatorname{argmax}}\;}
\newcommand{\la}{\left\langle}
\newcommand{\ra}{\right\rangle}

\newcommand{\be}{\begin{equation}}
\newcommand{\ee}{\end{equation}}
\newcommand{\ba}{\begin{align}}
\newcommand{\ea}{\end{align}}
\newcommand{\bea}{\begin{eqnarray}}
\newcommand{\eea}{\end{eqnarray}}

\usepackage{amsmath}

\newcommand{\va}{{\boldsymbol{a}}}
\newcommand{\vb}{{\boldsymbol{b}}}

\newcommand{\vh}{{\boldsymbol{h}}}

\newcommand{\vo}{{\boldsymbol{o}}}

\newcommand{\vr}{{\mathbf{r}}}
\newcommand{\vs}{{\boldsymbol{s}}}

\newcommand{\vw}{{\boldsymbol{w}}}

\newcommand{\vz}{{\mathbf{z}}}
\newcommand{\vT}{{\mathbf{T}}}

\newcommand{\vM}{{\boldsymbol{M}}}

\newcommand{\veta}{{\boldsymbol{\eta}}}
\newcommand{\vtheta}{{\boldsymbol{\theta}}}

\newcommand{\vmu}{{\boldsymbol{\mu}}}
\newcommand{\vnu}{{\boldsymbol{\nu}}}

\newcommand{\vphi}{{\boldsymbol{\phi}}}

\newcommand{\const}{{\rm const}}

\newfloat{Box}{tbp}{lob}
\definecolor{boxbg}{rgb}{0.9,0.9,0.9}
\newcounter{theoryboxc}
\newsavebox{\savetheorybox}
\newcommand{\printtheorybox}[1]{\fcolorbox{black}{boxbg}{#1}}
\newenvironment{theorybox}[1][]%
{\begin{Box}\begin{lrbox}{\savetheorybox}%
 \begin{minipage}{0.985\linewidth}
 \refstepcounter{theoryboxc}%
 \textbf{Box \thetheoryboxc. #1} %
}
{\end{minipage}%
 \end{lrbox}%
 \printtheorybox{\usebox{\savetheorybox}}%
 \end{Box}
}

\usepackage{lineno}

\title{How does the brain compute with probabilities?}

\author{
    Ralf M. Haefner*\\
    Department of Brain and Cognitive Sciences\\
    University of Rochester\\
    Rochester, NY\\
    \texttt{ralf.haefner@rochester.edu}\\
\And
    Jeff Beck*\\
    Department of Neurobiology\\
    Duke University\\
    Durham, NC\\
    \texttt{jeff.beck@duke.edu}\\
\And
    Cristina Savin*\\
    Departments of Neural Science and Data Science\\
    New York University\\
    New York, NY\\
    \texttt{csavin@nyu.edu}\\
\And
    Mehrdad Salmasi\\
    Gatsby Computational Neuroscience Unit\\
    Max Planck UCL Centre for Computational\\ Psychiatry and Ageing Research\\
    University College London\\
    \texttt{m.salmasi@ucl.ac.uk}\\
\And
    Xaq Pitkow*\\
    Neuroscience Institute\\
    Department of Machine Learning\\
    Carnegie Mellon University\\
    Pittsburgh, PA\\
    \vspace{.1mm}\\
    Department of Neuroscience\\
    Center for Neuroscience and Artificial Intelligence\\
    Baylor College of Medicine\\
    Houston, TX\\
    \vspace{.1mm}\\
    Department of Electrical and Computer Engineering\\
    Department of Computer Science\\
    Rice University\\
    Houston, TX\\
    \texttt{xaq@cmu.edu}\\
    ~\\
    {\bf *equal contribution}
}

\begin{document}
\maketitle

\begin{abstract}

This perspective piece is the result of a Generative Adversarial Collaboration (GAC) tackling the question `How does neural activity represent probability distributions?'. We have addressed three major obstacles to progress on answering this question: first, we provide a unified language for defining competing hypotheses. Second, we explain the fundamentals of three prominent proposals for probabilistic computations -- Probabilistic Population Codes (PPCs), Distributed Distributional Codes (DDCs), and Neural Sampling Codes (NSCs) -- and describe similarities and differences in that common language. Third, we review key empirical data previously taken as evidence for at least one of these proposal, and describe how it may or may not be explainable by alternative proposals. Finally, we describe some key challenges in resolving the debate, and propose potential directions to address them through a combination of theory and experiments. 
\end{abstract}


\section{Introduction}

Helmholtz observed that the sensory inputs to the brain are insufficient to give rise to the rich perceptual world that we experience, and that perception should be conceptualized as an active inference process which combines prior experiences with sensory inputs to form beliefs about behaviorally relevant features of the external world \citep{Helmholtz1867,kersten2003bayesian}. Over the past few decades, a large body of research has supported this view, demonstrating that humans and animals display behaviors that are sensitive to the relative uncertainties in their inputs and prior knowledge \citep{ma2012organizing}. Despite this extensive body of task specific research, there is still no consensus regarding either the means by which the computations that underlie this process are performed by neural circuits or the means by which neural activity represents uncertain beliefs\citep{fiser2010statistically,pouget2013probabilistic}. 

In the absence of computational limitations, the optimal way to deal with uncertainty is via Bayesian inference, which provides normative means by which subjective probabilities should be updated and utilized \citep{laplace1812theorie,jaynes2003probability}. However, debates continue about whether inference in the brain is actually probabilistic \citep{rahnev2020perception,orhan2017efficient}, and the jury is still out whether it is close enough to optimal to make Bayesian inference a useful mathematical framework for understanding the brain 
\citep{ma2012organizing}. 
Even within the context of Bayesian inference it is unclear whether probabilistic beliefs about inferred (latent) variables are computed `constitutively' across all latents, or are constructed `opportunistically' in response to task demands \citep{koblinger2021representations}. For the purposes of this paper we will assume that there is some population of sensory neurons whose activity can be interpreted as a probabilistic belief about some latent variable, and we will focus on the relationship between this belief and neural activity.

A series of prior studies has addressed this question and presented models falling into three broad categories of `neural codes': probabilistic population codes (PPCs) \citep{ma2006bayesian,jazayeri2006optimal,deneve2008bayesian,deneve2008learning,beck2008probabilistic,beck2012not}, distributed distributional codes (DDCs) \citep{sahani2003doubly,vertes2018flexible,pitkow2012compressive}, and neural sampling codes (NSCs) \citep{hoyer2003interpreting,fiser2010statistically,savin2014spatio,orban2016neural,haefner2016perceptual,aitchison2016hamiltonian}. While each of these models is supported by empirical evidence, it is often unclear how well the presented data exclude alternative models, and studies that directly compare multiple coding schemes are rare  \citep{grabska2013demixing,ujfalussy2022sampling}. Furthermore, comparisons across multiple papers are complicated by the fact that notation often differs, and differing assumptions are at times left implicit. In fact, recent work has identified some differences in assumptions and close relationships between models previously seen as mutually exclusive \citep{shivkumar2018probabilistic,lange2023bayesian}, pointing to a need for a systematic comparison of approaches, standard notation, and shared metrics of success. This review is an attempt to develop a common language and notation by proponents of different theories of probabilistic representations. After we present a unified and consistent language for representations (Section \ref{sec:abstractRepresentation}) and computations with them (Section \ref{sec:dynamics}), we then use that language to review the basics of the three major classes of theories, as well as formal connections between them and a case study of how they compare on a simple inference task (Section \ref{sec:models}). In Section \ref{sec:interpretationOfNeuralData} we systematically describe how these models each interpret a common set of empirical observations. Finally, in Section \ref{sec:future} we will note some of the inherent difficulties in comparing probabilistic coding schemes, and provide guidance for theoretical and empirical research designed to distinguish between the different theories.

\subsection{Why probability in the brain? 
}

The benefits of probabilistic computation are uncontroversial. It has been known since Laplace \citep{laplace1820theorie} and from the Dutch Book Theorem \citep{ramsey1926truth} that probabilities provide the optimal way to empirically reason about the world and form decisions in the presence of uncertainty. Furthermore, ample behavioral evidence has shown that perceptual and sensorimotor decisions are sensitive to changing uncertainty in a manner approximately consistent with probabilistic inference \citep{knill1996perception,knill2004bayesian,ma2006bayesian,fiser2010statistically,pouget2013probabilistic}. This implies that the brain represents uncertainty (if not entire probability distributions) over task relevant stimuli, implements the operations of probabilistic reasoning, and generates decisions based on that representation. We would like to understand the neural basis of these computations, and in particular whether there is a unifying theory that can explain how the brain computes with probabilities.

The existence of such a unifying theory is not a foregone conclusion. The brain may have learned to represent and manipulate probabilities in different ways for different tasks and variables it encounters. Probabilistic computations could arise in a highly flexible neural networks simply by extensive experience with naturally structured tasks \citep{orhan2017efficient}, as optimizing performance requires taking into account trial by trial fluctuations in uncertainty. Just because uncertainty must be represented and incorporated into the calculus of decision making does not mean that it is represented in the same generalizable manner for every task and latent variable, although there is some evidence in support of that claim \citep{houlsby2013cognitive}.
That said, several theories posit that there {\em are} general purpose, recurring motifs for representing and computing with probabilities. These motifs should arise with similar properties across a range of variables and tasks. If such a structure were to exist, then it would provide the brain with a powerful inductive bias that would generalize efficiently to new tasks. An inductive bias favoring learning and using probabilities could be embodied in large-scale architecture, microcircuit structure, and savvy plasticity rules \citep{sinz2019engineering}.

\subsection{What makes a `good' neural representation?}
Of course probabilistic information is already present at the retina, because one can always apply Bayes rule. That is not sufficient to count as a brain representation. Instead, a representation needs to be {\it used} in some way \citep{baker2021makes}.

Representations need to be evaluated in terms of how well they explain empirical observations.
However, it is also important to consider how well representations can be implemented and used by the brain. Three common desiderata for a `good' neural representation are: efficiency,  representational simplicity, and computational convenience (also see \cite{pohl2024desiderata}). Efficiency is typically measured in terms of bits per spike.  It makes use of the notion that one goal of the brain is to represent as much behaviorally relevant information as it can with minimal energy expenditure.  `Representational simplicity' refers to ease of decoding of a neural representation by downstream circuits constrained by computational complexity, time, and data. 

In statistical terms, a simple representation is one in which knowledge of low-order statistics like the mean and variance allow for efficient decoding. In contrast, a complex representation would relegate the encoding of objects and their poses, textures, and other properties, as well as the uncertainties about these properties, to complex, high-order statistics.  A related coding principle is computational convenience.  Certain representations make some computations easier to implement using the operations available to neural circuits.  It is typically assumed that linear operations are easy, even though individual biological neurons are capable of more complex computations \citep{poirazi2003pyramidal,beniaguev2021single,jones2021might,gidon2020dendritic}. 
Two example fundamental computations of particular interest in probabilistic inference are the sum rule and product rule of probability. Later we will see that different theories of probabilistic representations give these operations different complexities.

\begin{table}[htbp]
    \centering
    \rowcolors{1}{white}{gray!12}
    \caption{Glossary of notation.}
    \label{tab:notation}
    \begin{tabular}{rlrl}
        \hline
        symbol & meaning & symbol & meaning\\
        \hline
        $\vs$ & World state & & 
        \\
        $\vo$ & Observation & &\\
        $\vz$ & Latent variable in brain & $
        p(\vo|\vz)p(\vz)$ & Generative model in brain\\
        $t$ & Time & $T(\vz)$ & Statistic\\
        $q(\cdot)$ & Approximate posterior & $\eta$ & Natural parameter\\
        $\mathbb{E}_{q(x|y)}[\cdot]$ & Expectation over $q(x|y)$ & $\mu$ & Expectation parameter \\
        $\vr$ & Neural responses & $U(s,a)$ & Utility\\
        $\nu$ & Nuisance variable (external noise) & $\xi$ & Internal neural variability (internal noise) \\
        $\vw$ & Synaptic weight & $a$ & Action \\
        \hline
    \end{tabular}
\end{table} 

\section{What does it mean for the brain to represent probabilities?}
\label{sec:abstractRepresentation}

We say that a neural activity pattern represents a probability distribution if there is a mapping between neural activity and probability distributions {\em and} if subsequent neural computations are consistent with the rules of probability and the proposed mapping \citep{luce1990foundations,zemel1998probabilistic,baker2021philosophical,lange2022task}. For further discussion and nuance on the nature of representations, see \citep{baker2021makes}.

In the Bayesian perspective, probabilities are subjective constructs. However, these subjective probabilities are grounded in a model of the data generation process and computations based upon that model.  We assume that the brain's model of the world is based on a generative model of its sensory inputs (for a discussion of the differences between modeling the brain in terms of discriminative vs.\ generative models see \citep{peters2024does}). This comprises a set of assumptions about the latent variables that generate or cause the animal's sensory inputs. A good generative model is useful because it can allow the brain to explain the sensory data by drawing inferences about those latent causes (analysis by synthesis \citep{kersten2003bayesian}). Exact inference is intractable in general, and there will be algorithmic shortcuts or implementation constraints that lead to inferences that are only approximations to posterior distributions obtained through Bayes' rule. Theories about probabilistic brain computations often include such approximations.

\begin{figure}
    \centering
    \includegraphics[width=\linewidth]{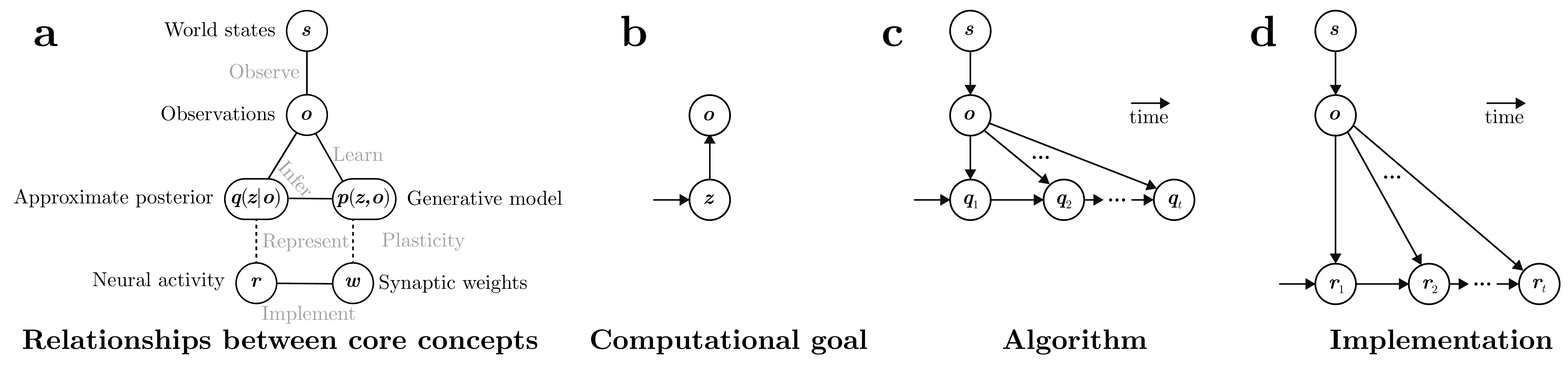}
    \caption{Relationships between key quantities for probabilistic inference. {\bf A}: Schematic of the different elements of Bayesian inference and its neural implementation. Not to be interpreted as a graphical model! {\bf B}: The computational goal of a Bayesian brain is to infer the brain's latent variables $\vz$ from observations $\vo$. {\bf C}: Inferential dynamics at the algorithmic level, for a static problem. Latent causes in the world generate observations, which the brain interprets through approximate inference dynamics in terms of its own latents $\vz$, eventually producing an action or decision. In theoretical models of inference, neural activity is a consequence of these algorithmic dynamics. {\bf D}: In reality, the physical mechanism or implementation of this process has a different causal diagram without the interpretable approximate posteriors, and the inferential dynamics are merely an abstract interpretation of the activity in the biophysical system.}
    \label{fig:schematic}
\end{figure}

To construct a testable, quantitative theory of how the brain computes with probabilities, we need to relate probabilities and brain signals, defined based on several considerations: 
First, what latent variables or events $\vz$ are the probabilities about? Second, what generative model $p(\vo|\vz)p(\vz)$ are the probabilities based on? 
Third, what aspects of neural activity $\vr$ represent the information about $\vz$?
Fourth, given the latent variables $\vz$ and observations $\vo$, what aspects of a posterior probability $p(\vz|\vo)$, or its approximation $q(\vz|\vo)$, are captured by $\vr$? 
Given answers to those four questions, we can construct a model for how $\vr$ represents $q(\vz|\vo)$ (Figure \ref{fig:schematic}).

\subsection{What is  \texorpdfstring{$\vz$}{z}? What variables are the probabilities about?}

A crucial question for neural theories of probabilistic computation is, what are the intermediate latent variables $\vz$ whose probabilities the brain may represent? These intermediate variables may have some causal status, {\it e.g.} the depth of different objects in a scene affects occlusion and thus the  visual input, or they may be pragmatic constructs that make it easier for a brain to summarize sensory sensory input in a way that is useful for predicting the effects of actions on future inputs.  For example, a set of oriented edges can be summed to create a compressed representation of an image while preserving the information needed to identify the pictured object \citep{olshausen1997sparse}. Either way, the brain has no access to an objective truth about the identities or values of intermediate latent variables and must rely on observations and the generative model assumptions about the relationships between latent variables to reason about them.

One possibility is that the brain should represent posteriors only over the variables that one can act upon. This is based on the idea that the goal of perception is, ultimately, to guide actions \citep{gibson1979ecological,shadlen2008intentional}. To select good actions you need predictions about what those actions will do. Moreover, since the future is unknown and different outcomes have different consequences, it is useful to represent probabilities of those outcomes in order to evaluate the benefits and risks of different options as is done in Bayesian reinforcement learning \citep{dayan2008decision,maloney2009bayesian} and active inference \citep{sajid2021active}. In this setting, variables that have no predictive power in the relevant action space do not influence behavior and thus need not be encoded. We call task-irrelevant variables `nuisance variables', $\vnu$, and a major computational goal of the Bayesian brain is to construct a representation of probabilities that are invariant to these nuisance variables.

An alternative possibility is that the brain constructs a task-independent model of the world that accommodates many situations, including those never seen before, and that the brain is always performing inference unconsciously, even when it is not performing a specific task (\citep{Helmholtz1867}, for a review see \citep{koblinger2021representations}).

However, even if inferring actionable variables is the eventual goal of the brain, there is often a complicated causal path to sensory observations from these latent variables. To perform inferences about the target variables, it may help to construct probability distributions over intermediate latent variables on that causal path \citep{peters2017elements}. Additionally, when tasks are amorphous and changing, it may help to represent latent variables that could later become actionable \citep{flesch2018comparing}, i.e. variables that lead to good generalization. Thirdly, the brain may benefit from constructing representations that facilitate subsequent learning or compress previous data. In each of these cases, representing a joint probability distribution about intermediate latent variables leads to better inferences.

\subsection{What is  \texorpdfstring{$\vo$}{o}? What evidence are the probabilities conditioned on?}

Probabilistic computations are based on evidence, whether it is the most immediate sensory observations or the evolutionary history of our ancestors. We usually assume that information from our evolutionary history is summarized in the architecture of the brain and modeled by prior belief about how the world works.  Information about what we have learned from previous experience is summarized in our synapses, while the neural activity (and perhaps short-term synaptic state \citep{mongillo2008synaptic}) is responsible for encoding information about recent sensory evidence and relevant latent variables. This recent evidence is what we will call observations $\vo$, and they determine the posterior probabilities $p(\vz|\vo)$ over latent variables that are of immediate interest. Below we discuss how the brain may approximate this ideal posterior by some other distribution $q$. A model of probabilistic computation should therefore specify what observations $\vo$ these probabilities are conditioned on.

What we count as an observation depends on the system we consider. Patterns of light are observations for the visual system, and patterns of sound are observations for the auditory system. But even within one modality, different subsystems receive different inputs: we might consider light as an observation for the retina, while the retinal ganglion cells' outputs are observations for the brain. In a broad, colloquial sense, we can consider an observation to be any input to a designated system. At the same time, there is a narrower, more technical definition of observation when we are considering probabilistic computation: an observation $\vo$ is whatever the posterior $q(\vz|\vo)$ is conditioned on. This requires making explicit modeling choices. 

In vision, for example, one might consider $\vo$ to denote the image, or the photoreceptor activations which provide the only evidence about the image, or the retinal output: none of these receive cortical feedback and thus can be treated strictly as inputs to downstream computations. Any computations performed by the retina itself, between image and retinal output, could be either modeled as part of the generative model using intermediate latent variables or, alternatively, as a potentially dynamic sensor that is not necessarily Bayesian in any meaningful way, and whose output is modeled as the observation from the perspective of the rest of the brain.

Ultimately, building a Bayesian model of some system requires defining the boundary of the system.  In a Bayesian framework, system inputs constitute the observations, and the output constitute ``actions.''  When modeling a cortical circuit, actions could simply consist of the transmission of the represented posterior. More generally, actions influence future observations and can be treated as either latents (represented by corollary discharge) or as part of subsequent observations.  For example, in an active inference or Bayesian reinforcement learning setting, the goal is often to compute a posterior distribution over actions that maximize rewards. At the behavioral level, only one action is actually selected which, if directly observed, becomes part of the subsequent observations.

Other considerations that determine where observation ends and inference begins include timescales of the relevant behavior, feedback between sensory and cortical areas, and the effects of actions such as eye movement on future observations.  For example, one might formalize intended actions as latents that affect future observations (e.g. as corollary discharge), or include actions themselves as observations.  Because of this complexity, it is often best to adopt a systems view in which observations consist of the complete set of signals that drive the system of interest.  

Where does sensory observation end and inference begin? Here, for simplicity of definition, we'll assume that the observation is constant over one inference step (whatever that means).
However, this is usually not the case and things get messy.
This is a tricky question, and one we do not claim to answer. There are questions of timescale, feedback, interaction with the environment, or an inability to make a clear separation between changing $\vo$ and the inference given a single $\vo$.

One crucial distinction between observations $\vo$ on which probabilities are conditioned, and the neural activity $\vr$ that represent those probabilities, is that $\vr$ should summarize the relevant aspects of the recent past, whereas $\vo$ provides information only from the current moment in time. 
Finally, the probabilities could in general incorporate all available evidence, including the animal's own actions $\va_{1:t}$, which influence the world and thus the probability through $p(\vz|\vo,\va)$. One might formalize intended actions either as latents or as part of the observations via feedback from the motor plant (e.g. as corollary discharge).

\subsection{What happened to  \texorpdfstring{$s$}{s}? What about the experimentally defined task stimulus?}

The term ``stimulus'' often refers to either specific properties of observable inputs, as for gratings or random dot kinematograms \citep{hubel1962receptive,parker1998sense}, or the entire observable stimulus $\vo$, as in studies with naturalistic images or sounds. Neither of these quantities necessarily directly correspond to the latent variables $\vz$ that neurons represent, as we describe below.

Experimental stimuli $\vs$ are often parameters of the observable sensory input provided to the brain $\vo$. For example, videos may be built from oriented gratings or dots moving with some coherence in some direction.  These variables are only a subset of the variables that define the sensory input, $\vo$. 

Many neuroscience studies take a philosophically distinct approach to asking questions about representations. They ask not about latent variables $\vz$ obtained from a computation model, but instead focus on experimentally controlled or measured stimuli, $\vs$, like orientation \citep{hubel1962receptive} or spatial location \citep{sherrington1906observations,o1971hippocampus} that strongly modulate neural activity. 

Importantly, the latent causes $\vz$ in the brain's model of the world need not agree with the experimentally-defined stimuli $\vs$.  Typically, the experimentally controlled $\vs$ only have some correlation with the brain's latent $\vz$.
For this reason it's important that we search for latents $\vz$ for which neural activity forms computationally useful representations. 
This latent variable approach eschews intuitive definitions of stimuli, like orientation or objects, and provides a framework for discovering sophisticated latents that better describe neural activity that drives behavior (see Box \ref{box:z-s}). 

\begin{theorybox}[The importance of the distinction between $s$ and $\vz$.\label{box:z-s}]
The distinction between $\vz$ and $s$ is crucial and consequential when empirically testing the predictions of different probabilistic encoding schemes. Here, we provide three illustrations of the importance of this distinction:

\paragraph{Primary visual cortex (V1):} Due to the strong orientation-selectivity of V1 neurons \citep{hubel1962receptive}, a dominant account of area V1 has been that it `represents orientation' (among other things, like spatial frequency, binocular disparity, etc.) Equating $s$ (here orientation) with $\vz$ leads to the conclusion that V1 responses are incompatible with a sampling-based representation, since that would imply that higher stimulus contrast, i.e. higher certainty about the stimulus orientation, leads to narrower tuning curves. However, the tuning curves in V1 appear to be approximately contrast-invariant, i.e.\ scale multiplicatively with contrast, suggesting a falsification of the neural sampling hypothesis. If, on the other hand, one assumes that $\vz$ represents the intensity, or absence and or presences, of (e.g.\ Gabor-shaped) image patches at the receptive field location \citep{olshausen1997sparse,bornschein2013v1}, then this conclusion changes, and even a sampling-based representation predicts an approximately multiplicative scaling of orientation tuning curves with contrast. In particular, \citep{shivkumar2018probabilistic} showed that implementing neural sampling in such a sparse model appears like a probabilistic population code (PPC) when interpreted as a code over orientation ($s$).

\paragraph{Medial temporal area (MT):} \citeauthor{hoyer2003interpreting} were the first ones to suggest that neural responses could be interpreted as samples from a posterior distribution over latent variables (in their case intensity of localized patches learnt from natural images), and that their variability might reflect the uncertainty in the brain's beliefs \citep{hoyer2003interpreting}. However, an analysis of the responses of a large number of neurons in area MT did not find a higher response variability for stimuli whose velocity was more ambiguous — an apparent contradiction of the neural sampling hypothesis (unpublished, private communication by Eero Simoncelli). The assumption underlying these analyses was that since MT neurons are strongly tuned to motion direction, $\vs$ (in addition to other variables), the represented beliefs would also be over $s$, such that higher uncertainty about $\vs$ should be reflected in higher response variability. However, it is not clear whether MT responses in fact represent beliefs over motion direction or velocity, not some other variable $\vz$, e.g. the absence or presence of motion primitives, which might make different predictions (in analogy to V1, see paragraph above).

\paragraph{Hippocampus (CA1):} In a recent study, \citeauthor{ujfalussy2022sampling} directly compared a set of neural predictions from three neural codes, PPC, DDC, and neural sampling, using simultaneous recordings from dorsal CA1 in the hippocampus. The analysis exploits the fact that neural responses during different phases of the theta cycle are believed to represent an animal's location at different times in past and future \citep{ego2007spatial}, with predictions increasing in uncertainty the further one goes into the future. This allows for strong qualitative predictions while avoiding the need to commit to a particular generative model about how location beliefs are updated in the light of sensory evidence. However, the analysis still relies on the assumption that the $\vz$ about which activity in area CA1 represents beliefs is the animal's trajectory. While this assumption is at least as plausible as the assumption that V1 represents orientation, or MT represents motion direction, there is ample evidence that CA1 represents more a general variable than location \citep{eichenbaum2018versus,jarzebowski2022different,sugar2019episodic}, and it is unclear whether the conclusions of \citep{ujfalussy2022sampling} would generalize for a broader definition of the latents.

\end{theorybox}

\subsection{What is \texorpdfstring{$p$}? What is the posterior probability that the brain would like to infer?}

Bayesian inference computes beliefs about unobserved latent variables $z$ given a set of observations $o$ and a generative model, $p(\vo|\vz)p(\vz)$, by applying Bayes' rule:
\begin{align}
    p(\vz|\vo) \propto p(\vo|\vz)p(\vz)
\end{align}
for a likelihood $p(\vo|\vz)$ and a prior $p(\vz)$. The likelihood captures the brain's assumptions about how observations $\vo$ arise from latent causes $\vz$. 
The prior captures the knowledge about the frequencies or values of causes and their dependencies on each other. This distribution is highly structured for natural inputs, and is often modeled as a hierarchical graphical model that efficiently expresses the conditional dependencies in $p(\vz)$. Together, these terms define a generative model of the world --- a way of explaining how the observations were generated by latent causes in the environment.
It is important to note that there need not be a direct correspondence between the brain's latents and quantities in the external world. As such, $p(\vo,\vz)$, denotes the brain's \textit{subjective} generative model of the world and $p(\vz|\vo)$ denotes the posterior consistent with that generative model, and not some unknown (and unknowable) probability that describes how the world actually works, e.g. in terms of physics.

\subsection{What is  \texorpdfstring{$q(\vz|\vo)$}{q(z|o)}? What is the brain's approximate inference?}

Inference in the brain is necessarily approximate. We denote by $q(\vz|\vo)$ as the brain's approximation to the exact posterior $p(\vz|\vo)$ given its own subjective generative model (here, assumed to be fixed after learning). The general underlying assumption is that the inference dynamics try to compute a $q$ that best approximates the desired $p$ by some measure.
While the nature of $q$ depends on the specific approximate inference algorithm, it should be a well-defined probability distribution and not a point estimate.

Note that there is some flexibility about which approximations define the inference $q$ versus the generative model $p$: one might define an altered generative model such that the $q$ is an exact posterior according to that model. Making assumptions explicit and testing how well they generalize is the key to discriminating between such model components.

\subsection{What is  \texorpdfstring{$\vr$}{r}? Which neural properties do the representing?}

In this paper, we will primarily consider neural activity as the seat of probabilistic computation. Previously proposed candidates in this context are membrane potentials \citep{orban2016neural}, spikes \citep{buesing2011neural,pecevski2011probabilistic,legenstein2014ensembles,savin2014spatio}, and spike rates \citep{hoyer2003interpreting,ma2006bayesian,vasudeva2016inference}. 
Some theories posit that probabilities are represented as a spatial code of spike counts in a long temporal window, manifested across neurons \citep{ma2006bayesian,vertes2018flexible}. Other theories including temporal sampling and timing codes \citep{hoyer2003interpreting,berkes2011spontaneous,orban2016neural,savin2014spatio} assert that the time series is the locus of probabilistic representations. Most generally, patterns of neural activity across both space and time may represent probabilistic information \citep{savin2014spatio}.

It is an open empirical question which of these types of neural response properties provide the most parsimonious description of probabilistic neural computations. Since spike times can be seen as a summary statistic of the underlying membrane potentials, and spike rates of the underlying spike times, a key question will be whether the respective lower level of description will have predictive power beyond that provided by the higher level description \citep{hoel2013quantifying}.

\subsection{What is the relationship between  \texorpdfstring{$\vr$}{r} and  \texorpdfstring{$q$}{q}? How does neural activity represent probabilities?}\label{sec:link-r-q}

\begin{figure}
    \centering
    \includegraphics[width=0.9\linewidth]{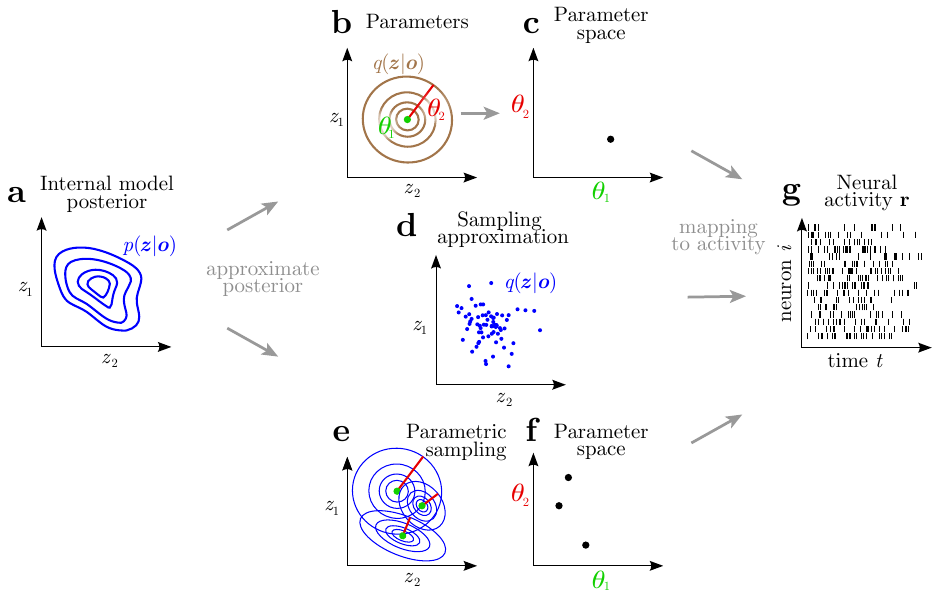}
    \caption{Examples of how a posterior distribution $q(\vz|\vo)$ ({\bf a}) could be mapped to neural activity. The distribution could be parameterized ({\bf b}) and these parameters ({\bf c}) could be mapped to neural activity ({\bf e}). Alternatively, samples from the posterior ({\bf d}) could determine neural responses. It is also possible to interpolate between these options, sampling the parameters \citep{lange2022interpolating} (Adapted from \citep{lange2022task}.) }
    \label{fig:QtoR}
\end{figure}

Neural activity evolves according to biophysical mechanisms. Probabilistic models propose that these mechanisms can be interpreted \emph{as if} they are implementing meaningful computations. The hypothesized link between $\vr$ and $q$ specifies the relationship between some biophysical properties and computationally meaningful ones (e.g. parameters of $q$, or samples from it, Figure \ref{fig:QtoR}). 
This link determines whether neural representations are mixed (multiple parameters or samples or statistics of $q$ contributing to each neuron's responses) \citep{rigotti2013importance} or `pure' (with only one aspect of $q$ contributing to each neuron) \citep{hoyer2003interpreting,fiser2010statistically}. And it determines whether $q$ is represented by individual neurons, or distributed across populations.   
Specifying this link is crucial for making testable neurophysiological prediction from any computational theory. Indeed, the mathematical link between $\vr$ and $q$ constitutes a primary distinction between the various probabilistic representational schemes.

Some coding schemes assume that the map between $\vr$ and $q$ is stochastic, even when the observation $\vo$ is fixed.  Trivially, this occurs when neural dynamics are noisy or not fully observed (nuisance variability), but can also be a product of a inference algorithm that that relies on optimization via stochastic gradient descent.  These kinds of nuisance and computational variability are expected to be present even when parameteric schemes are utilized.  Sampling based schemes, on the other hand, assume that neural variability is part of the approximate $q$ itself, i.e. neural variability is proportional to posterior's uncertainty \citep{lange2022task}.  
From an encoding perspective, the posterior is ``located'' in the circuit and sensory input that together generates stochastic responses. On the other hand, downstream circuits do not have access to this process, and only can access the stochastic neural activity. From a decoding perspective, therefore, we can say there may be a {\em distribution} over posterior distributions, and write any realized posterior as a sample from it, $q(\vz|\vo)\sim P[q(\vz|\vo)]$. 
This stochasticity has important implications for how we interpret neural variability, a topic we will return to in section \ref{sec:variability}.

\section{What are the dynamics of probabilistic computations?}
\label{sec:dynamics}

To understand how the brain computes with probabilities, we need to relate the dynamics of the neural activities $d\vr/dt$ to the dynamics of the posterior probabilities $dq/dt$ they represent. The specific relationship between these two quantities depends both on the format of the probabilistic representation and on the inference algorithm used to update approximate posteriors.  

Additionally, on a slower timescale, we assume that the circuits' parameters changes over time as $d\theta/dt$ as the circuit learns an improved generative model through $dp/dt$ and an improved approximate inference model. The next sections describe crucial properties of each of these computational dynamics.

\begin{figure}
    \centering
    \includegraphics[width=.45\linewidth]{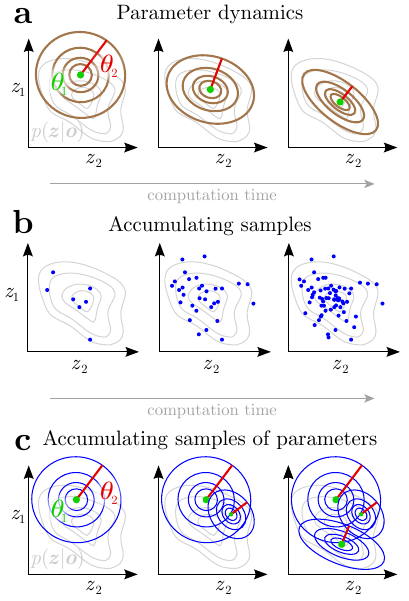}
    \caption{For a static posterior, $p(\vz|\vo)$, the approximate posteriors, $q_t(\vz|\vo)$ generally change over time. {\bf a}: Parametric codes allow parameters $\vtheta_t$ to depend on time. {\bf b}: Dynamics can produce sequences of samples, which when accumulated gradually fill out the posterior. {\bf c}: Dynamics can also produce sequences of sampled parameters \citep{lange2022interpolating} rather than samples of latent variables.}
    \label{fig:qdot}
\end{figure}

\subsection{What are the relevant timescales?}

Computations unfold over time. For a Bayesian brain, there are multiple temporal processes that are helpful to distinguish.
\textit{Learning:} At the slowest timescale, the brain learns the rules of the world. This corresponds to changes in the generative model, $\frac{d}{dt}p(\vz,\vo)$.
\textit{Inference:} This generative model describes a world of dynamic latent variables $\vz_t$ that can be inferred using a time series of observations $\vo_t$. These observations yield dynamic posteriors, $\tfrac{d}{dt}p(\vz_t|\vo_{-\infty:t})$, changing over time either as new evidence comes in, or as the brain anticipates future states. These posteriors may be dynamic even for an ideal Bayesian computation with unbounded computational abilities, since the dynamics are imposed by the time-dependent observations.
\textit{Algorithm:} At a faster timescale, an approximate Bayesian brain will use computations that unfold over time. Typically, those computations are iterative, and converge to the desired endpoint, through dynamics $\tfrac{d}{dt}q_t(\vz|\vo)$.
\footnote{We define $q\equiv q_t$ as the distribution implied by the current approximate computations, e.g. by the finite number of samples that have been generated, or by the current values of the parameters of $q$ even if they keep changing as part of an iterative algorithm.}
 \textit{Implementation:} Since a Bayesian brain requires a mapping between approximate posterior $q$ and neural activity $\vr$, a dynamic algorithm will manifest as neural dynamics $\tfrac{d}{dt}\vr$ even for a constant $q$. 

For simplicity, we will assume that the targeted posterior $p(\vz)$ is constant over time, using evidence $\vo$ that is also constant over time, so that neural dynamics correspond only to the algorithm and its implementation. In general, this is an unrealistic assumption as natural tasks invariably involve dynamic latent variables $\vz_t$ and series of observations $\vo_t$. Nonetheless, the restriction to static cases is useful here because it allows us to more easily distinguish different model types. The concepts we describe in this paper can be extended to inference over a dynamic world.

\subsection{What is  \texorpdfstring{$\dot{q}$}{dq}? What are the inference dynamics?}

For a fixed observation and a fixed posterior $p(\vz|\vo)$, the computation of an approximate $q$ unfolds over time, hopefully bringing $q$ closer to $p$ (Fig.\ref{fig:qdot}).
In temporal or spatiotemporal codes, the approximate posterior $q$ only manifests as a time series, so subsequent computations need to synthesize information across time. In contrast, spatial codes represent the $q$ completely within each relevant time window. 

For parametric codes and computations, the changes in the approximate posterior are captured by changes in the parameters, $q_t(\vz|\vo)=q(\vz|\vo,\eta_t)$, while for neural sampling codes they are the result of the changing set of samples.

\subsection{What is  \texorpdfstring{$\dot{p}$}{p}? How does the generative model change as it learns?}

As the brain gains experience in its environment, it can improve its model of the world, and use these changes to improve its inferences. Since we describe the brain's generative model of the world by $p(\vz,\vo)$, we refer to the learning-induced changes in that model as $\dot{p}$. Changes in an internal model are often attributed to synaptic plasticity $\dot{\vw}$, although there may be other physical mechanisms that may contribute, such as changes in bias or local nonlinearities, or even changes in the dynamics of short-term depression or facilitation.

\subsection{What is  \texorpdfstring{$\dot{r}$}{dr}? How do neural dynamics relate to computational dynamics?}

Biological mechanisms cause neural activities to evolve over time, determining the dynamics $\dot{\vr}$. In a Bayesian framework, these changes in $\vr$ are interpreted as changes in $q$ as it both incorporates new data and evolves toward the approximate posterior via the action of the inference algorithm employed.

Note that there may be some neural dynamics that are not relevant to the probabilistic representation, just as there may be some aspects of neural activity that do not encode relevant probabilities.

\section{Models of probabilistic representations}
\label{sec:models}

Now that we've introduced the core ingredients of probabilistic models of the brain, we will consider these ingredients for three major model classes: Probabilistic Population Codes (PPCs), Distributed Distributional Codes (DDCs), and Neural Sampling Codes (NSCs). Each has its own computational advantages and disadvantages, and the brain may incorporate elements of more than one model. In Section \ref{sec:interpretationOfNeuralData} we will review the empirical evidence in support of each.

\subsection{PPCs} \label{sec:PPCsIntro}

\paragraph{Key idea:} Probabilistic Population Codes (PPCs) assume that linear functions of neural activity represent {\it natural parameters} of a distribution, a core concept in probability we explain below. A direct consequence is that neural activity represents log probabilities. This makes it easy to multiply probabilities \citep{ma2006bayesian,rao2004bayesian}, as needed for cue combination and evidence integration. However, the other main probabilistic operation, marginalization, is more difficult.

\paragraph{What is $q$?}
In PPCs, neural activity represents exponential family probability distributions over latent variables $\vz$ by using simple encodings of natural parameters, $\veta$.  Natural parameters are a mathematically convenient way to parameterize a probability distribution: in $q(\vz)\propto \exp[\vphi(\vz)^\top\veta]$, the natural parameters are coefficients of sufficient statistics $\vphi(\vz)$ for that distribution. 
For any given parameterized distribution, there is a unique relationship between natural parameters and the expectations of the sufficient statistic. For example, in a Gaussian distribution, the natural parameters are the the inverse variance and the mean divided by the variance. 

\paragraph{What is $\vr$?}  As the name implies, PPCs are population codes, and authors typically assume that the relevant aspect of neural activity is firing rate or spike counts in large populations in some small time window.

\paragraph{Mapping between $r$ and $q$:}
A PPC assumes that the natural parameters, $\veta$, are linear functions of neural activity $\vr$: $\veta(\vo)=M\vr(\vo)$. This allows us to write the posterior distribution as
\begin{equation}
    q(\vz|\vr(\vo)) = \exp\left[\vphi(\vz)^\top M\vr(\vo) + \const(\vo)\right] 
    \label{eq:linearPPClog}
\end{equation}
\noindent where $(\phi(\vz)^\top M)_i$ represents the contribution of neural response $r_i$ to the posterior log probability over $\vz$. The basis functions $\vphi(\vz)$ determine the sufficient statistics of the associated exponential family distribution. For example, if $\vphi(\vz)$ is restricted to quadratic functions, then the associated posterior is a multivariate Gaussian.  

PPCs only describe the dimensions of neural activity that are relevant to encoding the posterior. Other orthogonal dimensions of $\vr$ are free to vary. (For downstream computations, these other dimensions serve as internal nuisance variables.) However, additional hypotheses about the neural activity may further constrain the connection between the task-relevant and -irrelevant aspects. For example, if $\vr$ represents spike counts, as for independent Poisson neurons, then these responses must be integers. This can constrain the particular posteriors that can be represented, and may also constrain otherwise task-irrelevant variations that ensure that spike counts are integers.  These additional assumptions are critical for making detailed neural predictions, to which we will return in Section \ref{sec:interpretationOfNeuralData}.

\paragraph{What is $\dot{q}$?}

In this paper, we focus on static inference problems, which can in principle be solved as a static nonlinear transformation of the sensory input, such as through a simple feedforward neural network with no dynamics. However, these problems may also be solved through an iterative algorithm \citep{vasudeva2016inference}. If the code remains consistent over time, then the neural dynamics would then correspond to iterative updates of the posterior $q$. Conversely, approximate inference schemes such as variational inference 
produce updates to natural parameters for the posterior, and therefore imply specific dynamics for the neural activity in the probabilistic coding dimensions \citep{beck2012not}.

\paragraph{What is $\dot{\vr}$?}
In PPCs, neural activity is linearly related to natural parameters, so when probabilities are multiplied, neural activity is added.  One consequence is that the amplitude of the neural response encodes confidence, with higher amplitudes corresponding to narrower posteriors. 
In contrast, marginalization is more difficult to perform on the neural activity, requiring non-linear operations such as coincidence detection and divisive normalization \citep{beck2011marginalization}.

\paragraph{What is $\vz$?} The PPC literature has largely focused on task-relevant latent variables in laboratory experiments performed by overtrained animals, such as orientation and contrast, or direction of motion and coherence. 
The initial focus on task relevant latents and decision variables has led to the misguided criticism that PPCs only applicable to simple tasks or are not fully Bayesian. However,
subsequent work showed how more general latent variables could be represented by PPCs \citep{beck2011marginalization,beck2012not, vasudeva2016inference} with network implementations that allow a variety of flexible computations on multivariate generative models.

\paragraph{What is $\vo$?} When investigating specific computations, $\vo$ is typically assumed to be either the sensory input or a pattern of neural activity that arises from the sensory periphery. For example, in vision, $\vo$ could be either the image itself, the photoreceptor absorptions, or the output of retinal ganglion cells. In an odor discrimination task, $\vo$ could be the activity of olfactory receptor neurons while the probabilistically encoding $\vr$ is the activity of downstream neurons in the olfactory bulb and piriform cortex.

\subsection{DDCs}\label{sec:DDCsIntro}
\begin{figure}[ht!]
	\centering
	\includegraphics[width =0.9\textwidth]{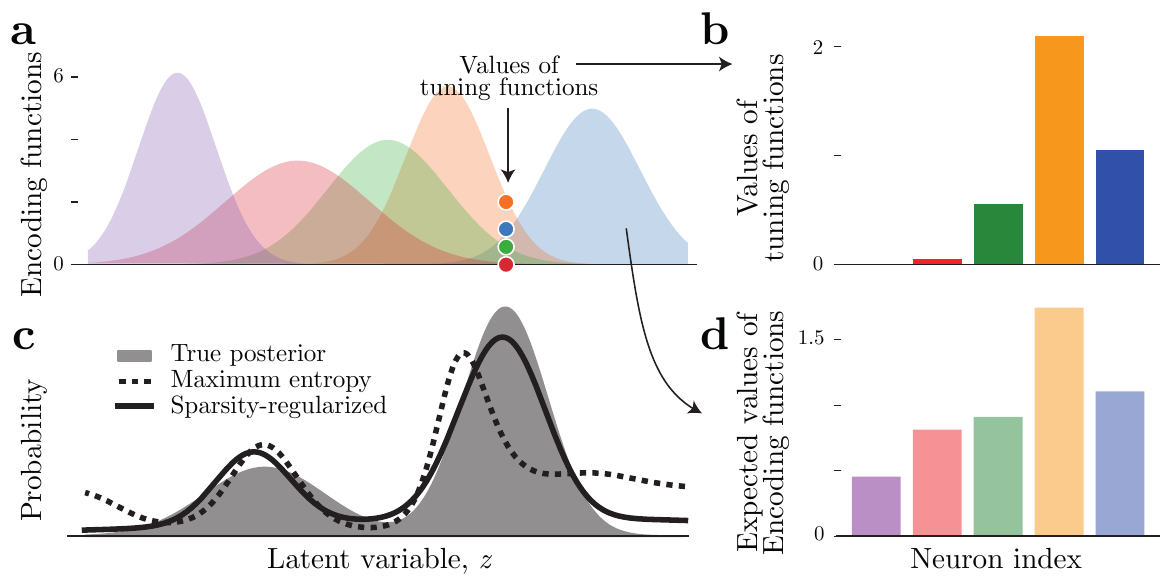}
	\caption{Distributed Distributional Codes (DDC): representation and decoding. (\textbf{a}) Five DDC encoding functions are assumed to represent the distribution. In the absence of uncertainty, e.g. $p(\vz|\vo) = \delta(\vz - \vz_\vo)$, the values of the encoding functions at $\vz_\vo$ represent the deterministic value of the latent variable (filled circles and the bar plot in {\bf b}). ({\bf c}) Under uncertainty, we illustrate the exact posterior distribution, $p(\vz|\vo)$, as a mixture of two Gaussian distributions (gray). The DDC representation is based on the expected values of encoding functions under the full posterior distribution ({\bf d}). The approximate posterior, $q(\vz|\vo)$, that is decoded from the representation depends on additional decoding choices, two of which are shown here: Dashed black line: the maximum entropy distribution derived from the DDC values in {\bf d}. Solid black line: the sparsity-regularized decoding of the belief.}
	\label{fig:Tuning_vs_DDC}
\end{figure}

\paragraph{Key idea:} 
In the absence of uncertainty, the brain can represent the deterministic value of the latent variable $\vz$ through a set of neuronal encoding functions $\{\phi_i(\vz)\}_{i=1}^K$ (Fig.\ref{fig:Tuning_vs_DDC}). Aligned with the conventional notion of tuning functions, the average firing rate of the neuron for the unknown value $\vz_0$ is given by $\mathbb{E} [\vr_i] = \phi_i(\vz_0)$, and some noise model (e.g. Poisson noise) captures the variability of the firing rate around the mean.
However, due to the noise in the sensory system and intrinsic epistemic uncertainty, the brain generally needs to deal with a distributional belief over the latent variable $\vz$, i.e. $p(\vz|\vo)$. Here, $p(\vz|\vo)$ refers to the exact posterior distribution in the generative model. 
A natural extension of the notion of tuning function is to assume that the firing rate of the neuron is determined by the weighted sum of the values of its tuning function at potential instances of the random variable $\vz$, where the weights correspond to the probability of the instances \citep{zemel1998probabilistic}. 

The firing rate of the neuron $i\in\{1,...,K\}$, is determined by
\begin{equation}
\vr_i  = \int_{\vz} \phi_i(\vz) p(\vz|\vo)d\vz.
\label{eq:DDC}
\end{equation}
The distributional belief $p(\vz|\vo)$ is therefore represented by $K$ expected values of the encoding functions, abbreviated here by the vector $\vr$ (Fig.\ref{fig:Tuning_vs_DDC}).

The codes of this scheme are referred to as Distributed Distributional Codes (DDC), as they provide a representation for "distributional" beliefs, and the representation is actualized through the "distributed" activity of neurons. 
DDC is a natural extension of tuning functions---in the absence of uncertainty, the posterior distribution is a Dirac delta function, $p(\vz|\vo) = \delta(\vz-\vz_\vo)$, and the DDC values are $\vr_i=\phi_i(\vz_\vo)$, $i=1,2,...,K$, which align with the definition of tuning functions.

Distributed distributional coding was first introduced in \citep{zemel1998probabilistic} as "extended Poisson model" to provide an encoding scheme for distributions. In addition to the encoding scheme, the decoding of the distribution is discussed, and it is argued that one can actually find a distribution over all the distributions that are consistent with the vector of firing rates $\vr$. Nevertheless, to simplify computations, an algorithm is suggested to approximate MAP estimate from the distribution over the distributions. The model was later renamed to "distributional population codes" in \citep{zemel1998distributional}, where it was used to explain the single-cell recordings and behavioral data in a multiple-motion task. The framework was subsequently extended to "doubly distributional population codes" to capture both uncertainty and multiplicity simultaneously \citep{sahani2003doubly}.

It should be noted that prior to the linear encoding approach in \citep{zemel1998probabilistic}, a linear decoding had been suggested in \citep{anderson1994basic, anderson1994neurobiological}. Instead of defining the encoding process, one can assume that the distributional belief $p(\vz|\vo)$ is a weighted sum of some basis functions $\psi_i(\vz)$. Therefore, the distributional belief can be computed through $p(\vz|\vo)  = \sum_i \rho_i \psi_i(\vz)$, where $\rho_i$ is the neuronal activity used to decode the distribution. In this approach, which is similar to kernel density estimation, the basis functions do not have direct relations with the tuning functions of the neurons. To find the decoding coefficients, $\rho_i$, different methods have been suggested including projection methods and EM algorithms. In the projection method, coefficients are computed by projecting the probability distribution on the basis functions, leading to a representation for $\rho_i$ which is similar to the encoding in \citep{zemel1998probabilistic}.

\paragraph{How it works:}
Probabilistic computation and inference frequently involves deriving the expected value of some function $f(\vz)$. One basic approach is to calculate $\mathbb{E}_{p(\vz|\vo)} [f(\vz)]$ is to decode the posterior distribution given the DDC values $\vr$, find an estimate of $p(\vz|\vo)$, denoted by $q(\vz|\vo)$, and then compute the expected value through $\mathbb{E}_{q(\vz|\vo)} [f(\vz)]$.
However, there is a simpler way. The primary characteristic of the DDC framework is its ability to compute expected values without needing to decode the probability distributions explicitly. Let $f(\vz) = \sum_i c_i\phi_i(\vz)$ represent a linear expansion of $f(\vz)$ with the encoding functions as the basis set. Then
\begin{align}
	\mathbb{E}_{p(\vz|\vo)} [f(\vz)] &= \sum_i c_i \mathbb{E}_{p(\vz|\vo)} [\phi_i(\vz)] \\
						   &= \sum_i c_i \vr_i
\end{align} 
Therefore, within the DDC framework, the expected value of an arbitrary function can be estimated by a weighted sum of the DDC values, significantly facilitating inference and learning procedures \citep{vertes2018flexible,vertes2019neurally,wenliang2020neurally}.

\paragraph{What is $\vo$?} 
In the DDC framework, the observation $\vo$ denotes the set of signals provided by the sensory system; alternatively, it signifies the output of the preceding processing stages that act as the "observation" for the subsequent layer. For example, in the early stages of visual processing, the retinal activity constructs the observation $\vo$, and is employed for inferring the DDC of posterior distribution over visual features, such as orientation and color. These inferred visual features can then be considered as "observations" in a generative model for higher-order processing, enabling another level of inference, like assessing the value of a shape in a decision-making task.
It is worth considering that in theory, one can envision a scenario where all inferences occur within an extensive hierarchical generative model. The sole observations in this context are the sensory signals, while the remaining variables constitute the latent variables requiring inference. Nevertheless, it appears that in order to maintain computational feasibility, the brain might employ separate generative models and propagate essential distributional beliefs among them.

\paragraph{What is $\vz$?} 
The latent variable $\vz$ encompasses any "unobserved" variable in the generative model that needs to be inferred. Within DDC, each latent variable is associated with a set of encoding functions whose expectations, with respect to the posterior distribution, determine the firing rates of the corresponding neurons. 
Latent variables in V1, for instance, may be linked to the orientation of image patches, whereas in CA1, they might denote the spatial location. The DDC encoding functions would then encode orientation in V1 and location in CA1. It is important to highlight that the concepts of orientation, location, and the like, as perceived by the observer, may not align precisely with the encoded information within the latent variables. The interpretation of the latent variables and establishing their connections with features in the external world poses a significant challenge. It is not yet fully understood how the brain dissects the stimuli to implant efficient features into the latent variables, and how an external observer can decrypt these features.

\paragraph{What is $\vr$?}
The neuronal firing rate is primarily defined by the expected value of the neuron's encoding function with respect to the posterior distribution. 
However, there are nuances that give rise to diverse extensions of this definition.
One might incorporate a noise model to account for the variability in the firing rate that is unexplainable by the posterior dynamics. For example, it can be a simple additive white Gaussian noise $\epsilon$, 
\begin{align}
	\vr = \mathbb{E}_{p(\vz|\vo)} [\phi(\vz)] + \epsilon,
\end{align}
or alternatively, the neuronal noise may manifest as Poisson noise, with an average given by $\mathbb{E} [\vr] = \mathbb{E}_{p(\vz|\vo)} [\phi(\vz)]$.

The other extension deals with the mapping between $p(\vz|\vo)$ and $\vr$ which can be non-linear, 
\begin{align}
	\vr = h \left( \mathbb{E}_{p(\vz|\vo)} [\phi(\vz)] \right) 
\end{align}
where $h$ is a non-linear function. Indeed, the combination of a non-linear mapping and a noise model would be a feasible approach for further extension.

\paragraph{What are $p$ and $q$?}
In a generative model with observation $\vo$ and the latent variable $\vz$, conducting exact inference yields the distributional belief $p(z|o)$.  
The DDC framework operates under the assumption that the brain's access is restricted to the DDC values $\vr$, with the flexibility for the approximate posterior $q(\vz)$ to be any probability distribution consistent with these DDC values. 
To decode the probability distribution, one needs to find the probability distributions $q(z)$ that satisfy the constraints $\vr_i = \int q(\vz)\phi_i(\vz) d\vz$, for $i=1,2,...,K$. This problem is known as generalized moment problem and have been studied extensively, initially for polynomial moments and later for generalized moments \citep{schmudgen2017moment,kemperman1968general}. 
A given DDC vector $\vr$ can correspond to none, one, or more than one distribution. When multiple distributions are associated with $\vr$, various optimization functionals can be utilized to find the "optimal" posterior $q^*(\vz|\vo)$. For a given utility functional $U$,  the optimal posterior is derived from
\begin{align}
	q^*(\vz|\vo) &= \argmax{q} U(q) \hspace{0.3cm} \text{subject to:}\\
				 & \int q(\vz)\phi_i(\vz) d\vz =\vr_i ,\hspace{0.2cm}  \text{for} \hspace{0.1cm} i\in\{1,2,...,K\} \\
				 &\int q(\vz) d\vz = 1 \\
				 & q(\vz)\geq 0
\end{align}
A well-known choice for the utility functional is the entropy of the probability distribution, $U(q) = -\int_{\vz} q(\vz)\log q(\vz)d \vz $. The maximum entropy solution of the generalized moment problem is an exponential family distribution \citep{wainwright2008graphical}.
\begin{align}
	q(\vz|\vo) = \exp\left(\sum_{i=1}^K \phi_i(\vz)\eta_i - A(\eta_1,\eta_2,...,\eta_K)\right),
\end{align}
where $A(.)$ is the log-partition function.
This approach yields the probability distribution with the highest uncertainty while satisfying the expectation constraints imposed by the DDC values. The natural parameters, $\eta$, correspond to the Lagrange multipliers used for solving the optimization problem. 

Alternative utility functionals include measures such as the sparsity of distribution in a given basis, the smoothness of the distribution, and others. In Fig. \ref{fig:Tuning_vs_DDC}, the maximum entropy distribution and sparsity-regularized distribution have been derived for the given DDC values.
Another proposition involves considering the set of distributions $q(\vz|\vo)$ that satisfy the expectation constraints and deriving a posterior distribution over all these distributions \citep{zemel1998probabilistic}. In fact, by assuming a sparsity measure for optimization and additive noise for the DDC values, one can use an empirical Bayes method \citep{ji2008bayesian} to derive the posterior distribution over all the beliefs that satisfy the DDC values \citep{salmasi2022learning}.

As previously mentioned, DDC endeavors to bypass the decoding of the distributions $q(\vz|\vo)$---instead, it transforms probabilistic inferences/computations into the determination of expected values of certain functions. These expectations are computed straightforwardly through weighted sums of the DDC values.

\paragraph{Mapping between $q$ and $\vr$:}
Any approximate posterior distribution $q(\vz|\vo)$ should satisfy the generalized moment constraints $\int q(\vz)\phi_i(\vz) d\vz =\vr_i$, for $i=1,2,...,K$, though the noise variance will allow some deviations from equality. The mapping from $\vr$ to $q$ is related to the generalized moment problem that was discussed in the previous part. To derive the maximum entropy distribution for the given DDC $\vr$, the natural parameters $\eta(\vr)$ are calculated from the set of non-linear equations, 
\begin{align}
	\int_z \phi_i(\vz)\exp\left(\sum_{i=1}^K \phi_i(\vz)\eta_i - A(\eta_1,\eta_2,...,\eta_K)\right) d\vz = \vr_i , \hspace{0.2cm} i\in\{1,2,....,K\} 
\end{align}
where
\begin{align}
	A(\eta_1,\eta_2,...,\eta_K) = \log \int \exp\left(\sum_{i=1}^K \phi_i(\vz)\eta_i\right)d\vz,
\end{align}
and a closed-form distribution is obtained for $q(\vz|\vo)$. However, as stated earlier, maximum entropy is not the sole approach to derive $q(\vz|\vo)$ from $\vr$. 

\paragraph{What is $\dot{r}$:}
Various algorithms have been suggested for conducting inference and learning in the DDC framework. 
Helmholtz machines propose an elegant method for joint inference and learning---the wake-sleep algorithm iteratively refines the generative model and recognition network of the Helmholtz machine, and concurrently, learns the generative model of the world and acquires the capacity to infer the latent variables.\citep{dayan1995helmholtz}.

DDC has been integrated into the Helmholtz machine to provide a biologically plausible mechanism for inference and learning \citep{vertes2018flexible}.
It is assumed that the generative model belongs to deep exponential family models---the conditional probabilities and priors are exponential families each characterized by specific sufficient statistics. 
The recognition network has a similar hierarchical structure and each layer is associated with a set of DDC encoding functions. The recognition network performs inference by mapping the observations to the DDC values of each layer, and the recognition outputs are interpreted as representing a posterior distribution with maximum entropy, meaning that the approximate posterior corresponds to an exponential family.

During the sleep phase, the generative model is used to generate samples of latent variables and sensory observations (dream sequence). 
The goal of the recognition network is to minimize the Kullback-Leibler divergence between the deep exponential family distribution of the generative model and the approximate maximum entropy distribution of the recognition network. Since both probability distributions are from exponential family, the parameters of the recognition network are modified to minimize the difference between the DDC values of the recognition network and the expectations of the sufficient statistics of the generative model. 

During the wake phase, sensory observations are collected and utilized by the recognition network for inferring the DDC values of the posterior distributions over the latent variables. Subsequently, the sensory observations and the DDC values are used to update the parameters of the generative model in order to increase the variational free energy. It is shown that the gradient of the free energy can be derived by calculating the expected value of some functions of the latent variables. This means that by using a linear expansion for these functions, the gradient of free energy can be approximated by weighted sums of the DDC values. The remaining issue is the learning of expansion coefficient that can be conducted using the generated samples in the sleep phase \citep{vertes2018flexible}.

\paragraph{What is $\dot{q}$:}
The dynamics of the approximate posterior $q$ is intricately linked to the dynamic evolution of the DDC values $\vr$.
In the Helmholtz machine, for example, the sensory observations and the current weights of the recognition network determine the DDC values of each layer. By adopting the maximum entropy distribution, the conditional distribution of each layer is mapped to an exponential family, whose sufficient statistics are the DDC encoding functions of that layer and the natural parameters are calculated by the given DDC values.
It is worth emphasizing that in the realm of DDC computations, there is a possibility of encountering DDC values that lack feasibility, meaning there is no corresponding distribution for them. Nevertheless, with a rich set of encoding functions, the probabilistic computations can still maintain a high degree of precision.

\paragraph{What is $\dot{p}$:}
Depending on the generative model employed, various algorithms can be used for learning the model's parameters. 
In the case of a deep exponential family model, the parameters of the generative model can be learnt by the wake-sleep algorithm, as discussed previously \citep{vertes2018flexible}. The natural parameters of each layer in the deep exponential family are calculated by a parametrized function of the parent variable. The parameters of these functions are updated by calculating the gradient of the variational free energy. The generative model is affected both in wake and sleep phases. The variational free energy is calculated through the expected values of some linear functions of sufficient statistics. The expectations are calculated by the weighted sums of the DDC values during the wake phase, and the weights of the linear functions (expansion coefficients) are learnt through the samples of the generative model in the course of the sleep phase.

\subsection{Neural sampling}\label{sec:SamplingIntro}

\paragraph{Key idea:}
A probability distribution can be approximated by a collection of samples from it, rather than by its parameters. The key idea underlying `neural sampling' is that the neural activity in small time bins can be interpreted as one or more samples from the brain's posterior, and that over time or space, the distribution of neural activity reflects the posterior \citep{hoyer2003interpreting,fiser2010statistically}.
To implement this, stochastic recurrent dynamics explore the state space of latent variables, occupying states in proportion to their posterior probability. Given these samples, the brain can directly estimate expectations of any function of the latent variables, which is helpful for choosing actions. For example, expectations can compute the posterior mean to generate a single estimate, a posterior variance to quantify uncertainty, or expected reward to compare states and actions.

\paragraph{What are $p$ and $q$?} Samples may be drawn either from the exact posterior $p$, or from an approximate posterior distribution, $q(\vz|o)$, as in stochastic variational inference \citep{savin2011two, hoffman2013stochastic}. 

\paragraph{What is $\vr$?} Neural sampling dynamics come in three main flavors, differing by which aspect of neural activity encodes the samples: (1) membrane potential (continuous $\vz$)\citep{orban2016neural,banyai2019stimulus}, (2) spike/no spike (binary $\vz$) \citep{buesing2011neural,pecevski2011probabilistic,haefner2016perceptual,shivkumar2018probabilistic}, or (3) firing rate (continuous latent $\vz$) \citep{hoyer2003interpreting,haefner2016perceptual,echeveste2020cortical}. 

\paragraph{What is the mapping between $q$ and $\vr$?} Most proposals assume a one-to-one map between latent dimensions and responses of individual neurons, although the two can also be related less directly via a linear map, $\vM$ \citep{savin2014spatio}. 
\begin{equation}
    q(\vz|\vo)=\frac{1}{n}\sum_{k=1}^n \delta(\vM\vr^{(k)}-\vz)
\end{equation}
where $n$ is the number of samples.

\paragraph{What is $\dot{q}$? } Most neural sampling proposals consider static inference problems, in which a posterior is inferred for a given stimulus. On that (shortest) timescale, $\dot{q}$ simply reflects the additional samples generated over time, successively refining the posterior approximation. The sampling idea can be expanded to time-varying inference problems in which the posterior evolves over time on the time scale of the stimulus dynamics, for instance by neural dynamics analogue of particle filtering \citep{lee2003hierarchical,kutschireiter2018particle}.

\paragraph{What is $\vz$?} Prior work on neural sampling has primarily focused on generative models of natural images, including linear Gaussian models and sparse variants \citep{olshausen1996emergence,olshausen1997sparse,hoyer2003interpreting,haefner2016perceptual,shivkumar2018probabilistic} or Gaussian scale mixtures  \citep{schwartz2001natural,wainwright2002,orban2016neural,banyai2019stimulus}. Other examples include a hierarchical extensions of these \citep{haefner2016perceptual,banyai2019stimulus,csikor2023top}, a sparse linear Poisson model of olfactory inputs \citep{grabska2013demixing,grabska2017probabilistic}, or a probabilistic formalization of memory retrieval where the latents are possible items retrieved from memory \citep{savin2011two,savin2014optimal}. In these works, latents are either continuous \citep{hoyer2003interpreting,grabska2013demixing,savin2014spatio,orban2016neural,haefner2016perceptual,banyai2019stimulus}, e.g.\ representing the intensity of an odorant or the amplitude of a Gabor feature in the visual input, or discrete \citep{buesing2011neural,savin2011two,savin2014optimal,haefner2016perceptual,shivkumar2018probabilistic}, e.g.\ binary variables for different task contexts.   

\paragraph{What is $\vo$?} The nature of the relevant observations depends on the generative model. They could be a retinal image or the whitened output of retina in most vision related studies, receptor activity in olfaction \citep{grabska2013demixing}, spikes in other brain regions \citep{beck2012not}, strength of synapses storing information about past experience in \citep{savin2011two,savin2014optimal}.

\begin{figure}
    \centering
    \includegraphics[width=\linewidth]{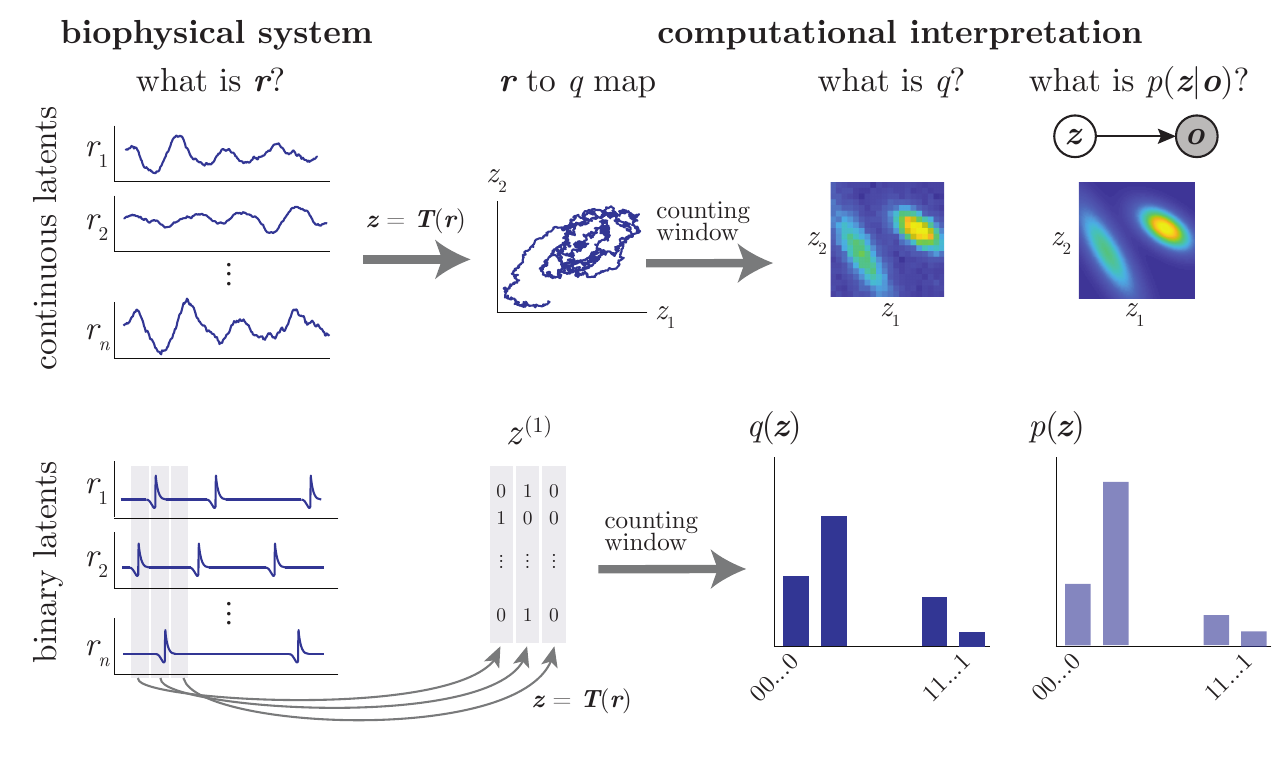}
    \caption{Illustration of neural sampling codes using both continuous and binary latents.}
    \label{fig:sampling}
\end{figure}

\paragraph{How it works:} The details of the circuits implementing neural sampling can differ across proposals, but they generally take the form of stochastic dynamics that map a current sample into a new one, via a transition probability. For continuous latent variables, the simplest example is Langevin sampling where the dynamics perform gradient descent on an `energy' (given by the negative log posterior), with additive gaussian noise:
\begin{align}
    \vz^{(t+1)} &= \vz^{(t)} + \alpha\nabla \log p(\vz|\vo) + \boldsymbol{\epsilon}^{(t)}
\end{align}
where $\nu$ determines the step size and $\epsilon$ is zero-mean Gaussian noise whose variance only depends on $\alpha$ (Figure \ref{fig:sampling}A). Another classic example algorithm is Gibbs sampling in which new samples are drawn one dimension of $\vz$ at a time: $z^{(t+1)}_k \sim p(z_k|\vz_{\neg k}=\vz_{\neg k}^{(t)})$ where $\vz_{\neg k}^{(t)})$ denotes the last sample, including all dimensions of $\vz$ other than the $k$th (Figure \ref{fig:sampling}B). Compellingly, the circuits resulting from Gibbs sampling only rely on local connectivity \citep{buesing2011neural,haefner2016perceptual}, matching classic anatomical results \citep{felleman1991distributed}. The accuracy of sampling-based approximate inference critically depends on time, which makes sampling speed a key consideration when assessing the computational efficiency of different sampling schemes. Many of the recent theoretical efforts have been motivated by the conjecture that sampling in the brain is likely accelerated compared to classic sampling algorithms like Gibbs sampling, for instance through the use of balanced amplification \citep{hennequin2014fast,echeveste2020cortical}, network oscillations \citep{savin2014optimal,aitchison2016hamiltonian}, or other biophysical features \citep{buesing2011neural}.

\subsection{Formal connections between the different proposals}

DDC and PPC are both categorized as parametric schemes, however, DDC is defined as an encoding framework, while PPC is primarily introduced as a decoding scheme \citep{lange2023bayesian}. 
Within the DDC framework, the posterior distribution $p(\vz|\vo)$ is "encoded" by the expected values of the DDC encoding functions, determining the neuronal activity $\vr$. In contrast, PPCs depart from defining the encoding scheme, and instead, assume that the logarithm of the posterior distribution can be approximately "decoded" through a weighted sum of a set of basis functions, where the weights correspond to the neuronal firing rates $\vr$. 
This decoding-based scheme can be compared to the linear decoding approach in \cite{anderson1994basic, anderson1994neurobiological}, where it is assumed that the posterior distribution can be approximated through a weighted sum of given basis functions (similar to kernel density estimation). The distinguishing factor in PPC is the application of linear decoding to the "logarithm" of the probability distribution.

By narrowing our scope to a specific decoding approach for DDC, namely the maximum-entropy decoding, we can strengthen the connection between PPC and DDCs. In this case, the approximate posteriors in both DDC and PPC are members of the exponential family---PPC represents the natural parameters of the exponential family by a linear mapping of neural activity, while DDC encodes the expectations of sufficient statistics (mean parameters). 
Therefore, under the premise of maximum-entropy decoding, PPCs and DDCs can be transformed into one another through the one-to-one correspondence between natural parameters and mean parameters in exponential family distributions, though the conversion can be computationally expensive.

For the remainder of this subsection, we will assume that the underlying posterior (or approximate posterior) is a member of the exponential family of distributions with finite dimensional sufficient statistic $\vT(\vz)$ and natural parameters $\veta$.  The distinction is that PPCs represent the natural parameters $\veta$ linearly in neural activity, whereas DDCs represent the expectation parameters $\vmu$ 
linearly in neural activity.  Because these two types of parameters can be uniquely related to each other, 
PPCs and DDCs are both equally expressive and can, in principle, be mapped onto each other. In fact, this conversion is a primary goal of many inference problems \citep{wainwright2008graphical}, such as inferring marginals (directly related to expectation parameters) from a joint distribution specified by natural parameters. 
In practice, however, exact conversion is often difficult or even intractable, so probabilistic reasoning requires sophisticated approximation schemes or recognition networks that learn to approximate the relationship between $\vmu_\vT$ and $\veta$.  Curiously, the principle means by which the relationship between natural parameters and expectations are discovered is via generating samples conditioned on the natural parameters and then using those samples to approximate the expectations.  

It is worth noting, however, that many inference algorithms function by iteratively updating expectations and natural parameters, suggesting that both DDCs and PPCs can interact fruitfully to perform fundamental computations. 
For example, consider belief updating using variational inference with a factorized posterior, $q(\vz_1,\vz_2)=q(\vz_1|\veta_1)q(\vz_2|\veta_2)$.  This algorithm uses iterative posterior updates that obey
$$\veta_1\cdot \vT(\vz_1) = \left<\log p(\vz_1,\vz_2)\right>_{q(\vz_2|\eta_2)} + {\rm const}$$
where $p(\vz_1,\vz_2)$ is the target posterior to be approximated by $q$.  When $\vT(\vz)$ is expressed as a set of orthonormal basis functions, the joint distribution of observations and latents can be written as a linear combination of outer products and thus 
\begin{align*}\veta_{1i} &= \sum_{j} a_{ij} \left< T_j(\vz_2) \right>\\
\veta_{2i} &= \sum_{j} b_{ij} \left< T_j(\vz_1) \right>
\end{align*}

A linear PPC assumes that the left hand sides of the above equations are linear in neural activity while a DDC assumes that the expectations on the right hand side are linear in neural activity.  Thus, in this setting, a linear PPC representation for $q(\vz_1)$ corresponds to a DDC representation of $q(\vz_2)$ and vice versa since

$$M_1^{\rm PPC}\vr_1^{\rm PPC}=\mathbf{A} \vM_2^{\rm DDC}\vr_2^{\rm DDC}$$
$$M_2^{\rm PPC}\vr_2^{\rm PPC}=\mathbf{B} \vM_1^{\rm DDC}\vr_1^{\rm DDC}$$

More generally it can be shown that in a multilayer generative model implemented using a DDC, a PPC for latent variables in layer $j$, can be constructed from a quadratic combination of neural activity representing a DDC in layers $j-1$ and $j+1$.  Indeed, the same quadratic combination of DDC representations is what drives learning in the DDC framework precisely because the quantities being learned parameters that are linearly related to natural parameters.  As a result, learning signals in a network in which neural activity forms a DDC, the learning signals form are PPC representations of the corresponding posteriors.

Because they are both based upon exponential family distributions and associated approximate inference schemes, the difference between PPCs and DDCs comes down to the empirical question of whether the brain uses activity to linearly represent natural parameters (PPCs) or expectation parameters (DDCs).  Computationally, these representational schemes differ in computational convienence and efficiency.  For example, the product rule (e.g. evidence integration) is linear in log-probability or in natural parameters, which makes them easy to implement for linear PPCs, but is nonlinear in probability.  Similarly, marginalization is a linear operation on probability and hence also linear for expectations so this operation is 'easy' for a DDC while the product rule requires a quadratic operation.  Of course, this argument assumes the linear operations are in some sense preferred by neural circuits.  

Inference generally requires both computations, and previous work has shown that a network capable of implementing a quadratic non-linearity and divisive normalization is sufficiently computationally expressive to implement both evidence integration and marginalization with a PPC \citep{beck2011marginalization}. In the Kalman filter, for example, all of the complexity of the standard equations are explained simply by the need to switch back and forth between a natural parameter representation to use the product rule of probability when updating posteriors with new evidence, and an expectation parameter representation to use the sum rule of probability when marginalizing over states. This switching suggests that, in the brain, one might expect to find signatures of both PPCs and DDCs at different stages or in different functional subpopulations.

Similar links exist between sampling and DDCs.  For instance, for generative models based upon exponential family distributions, NSCs require the evaluation of the sufficient statistic $\vT(\vz)$ or its gradient for each sample $\vz$.  Since a DDCs is present when the average of this quantity is available, a simple  average of neural activity associated with an NSC leads to a DDC.  

\subsubsection{Special cases and alternative proposals}

It is worth noting that the Free Energy Principle \citep{friston2010free} is a special case of a parametric code in which neural activity represents the parameters of $q$. In its instantiation as predictive coding, it further assumes a mean-field approximation to the full posterior, i.e. $q(\vz)=\prod_i q_i(z_i)$ in which each of the $q(z_i)$ is Gaussian \citep{gershman2019does}. Such a representation is extremely limited in its expressive power compared to more general PPCs, DDCs, or NSCs, since it cannot represent any dependencies in the posterior $p$ due the factorization.

Furthermore, PPCs, DDCs, and NSCs are not the only possibilities for probabilistic representations. For example, recent work in distributional reinforcement learning has proposed that the brain may use expectile codes \citep{dabney2020distributional}. Such representations could themselves be constructed based on other probabilistic codes, such as sampling \citep{rullan2021sampling}.

It is also possible to interpolate between the model classes. For example, it is possible to sample the parameters of a distribution, rather than sampling the latent variables directly, like a sampled mixture of PPCs or DDCs \citep{lange2022interpolating} (Fig.~\ref{fig:qdot}{\bf d--e}, Fig.~\ref{fig:QtoR}{\bf c}).

\begin{figure}
    \centering
    \includegraphics[width=\linewidth]{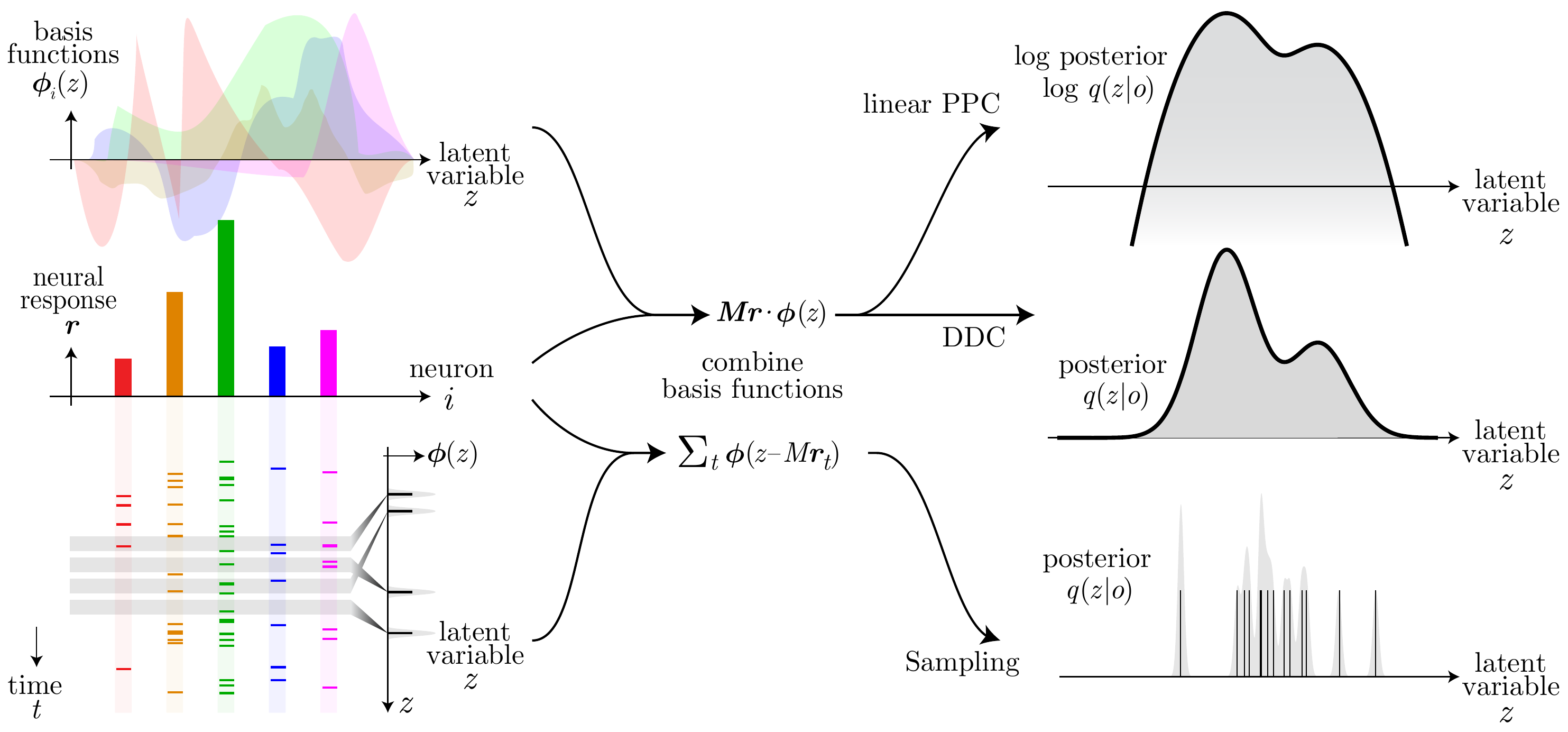}
    \caption{Illustration of the similarities and differences between linear PPCs, DDCs, and NSCs. In all three encoding schemes, the approximate posterior $q(z|o)$ can be written as a sum of basis functions $\phi_i(z)$. The essential difference between PPCs and DDCs on the one hand, and NSCs on the other, is that for PPCs \& DDCs the neural responses $\vr$ determine the \emph{amplitude} of each basis function, while for NSCs the neural responses determine the \emph{location} of each basis function. Another difference is that the number of basis functions for PPCs and DDCs is fixed, while for NSCs it is variable, increasing with time (one sample per time window).
    Note that while this illustration focuses on a scalar $z$, all three schemes also work for multidimensional $\vz$. The basis functions for DDCs depend on an additional constraint like maximum entropy, similarly to the conversion of a sum of $\delta-$functions into a histogram for sampling. To maximize similarity between the schemes this illustration uses a distributed NSC scheme where samples are linearly read out from neural responses; most existing NSCs assume a 1-1 relationship between a single neuron's $r_i$ and a corresponding $z_i$ (also see Fig.~\ref{fig:sampling}).} 
    \label{fig:PPCDDCSampling}
\end{figure}

\subsection{Case study: cue integration}

\begin{figure}
    \centering
    \includegraphics[width=.3\linewidth]{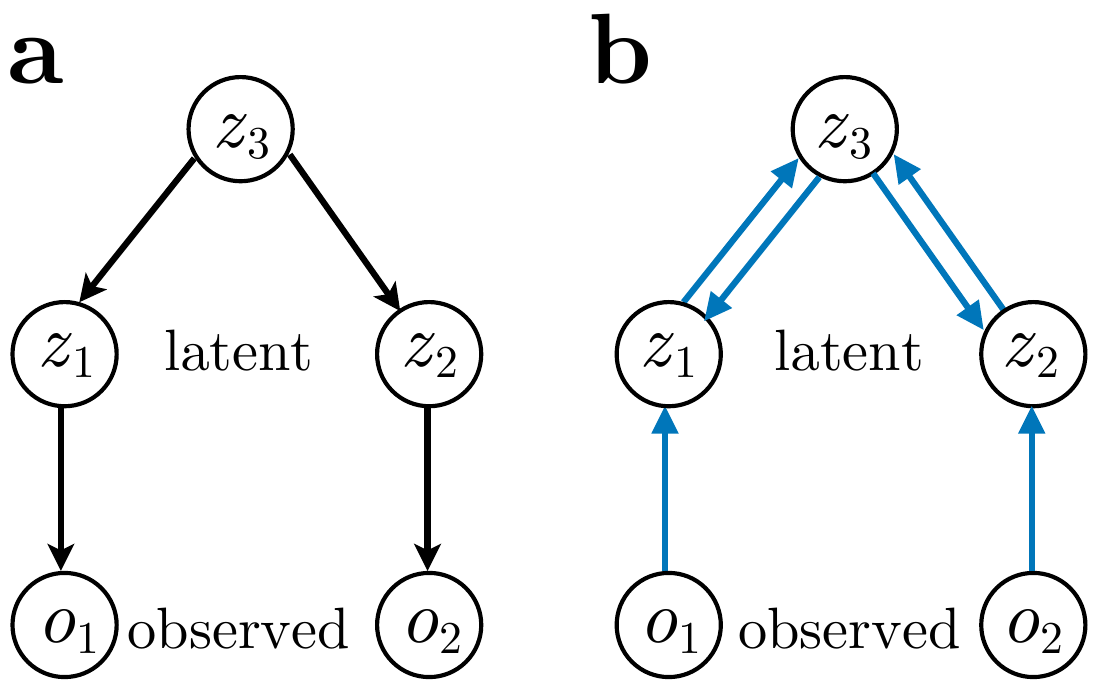}
    \caption{Illustration of probabilistic computations: cue integration: \textbf{(a)} Directed probabilistic graphical model. \textbf{(b)} Information flow during inference, i.e. the computation of $p(z_1,z_2,z_3|o_1,o_2)$.}
    \label{fig:cuecombination}
\end{figure}

Figure \ref{fig:cuecombination} shows a simple case study we will use to showcase computations needed for all three classes of theories. Here we use a probabilistic graphical model in which one top-level variable $z_3$ affects two lower-level latents $z_1$ and $z_2$, each generating its own observations $o_1$ and $o_2$ (Figure \ref{fig:cuecombination}). We assume that tasks based on this model depend only on single latent variables, so the goal is to calculate marginal probabilities conditioned on all of the observations. In other words, the goal for inference in this model is to compute a representation of the high-level marginal posterior $p(\vz_3|\vo_1,\vo_2)$, as well as the low-level posterior $p(\vz_1|\vo_1,\vo_2)$ that accounts for both direct evidence from $\vo_1\to\vz_1$ {\it and} indirect evidence $\vo_2\to\vz_1$. Inference in this model is a nice case study because it requires use of both the product rule of probability, when combining the direct and indirect evidence, and the sum rule, when marginalizing over latent variables.

\subsubsection{PPC}

Integrating independent cues is straightforward in PPCs: this operation corresponds to the product rule for probabilities, so if neural activity is proportional to log probability, this means simply adding neural activity. We could do this directly if we had separate PPCs $\vr^{(3|1)}$ and $\vr^{(3|2)}$ for distributions over the higher-level variable $\vz^3$, corresponding to indirect evidence $q(\vz_3|\vo_1)$ and $q(\vz_3|\vo_2)$. Then our final PPC for $q(\vz_3|\vo_1,\vo_2)$ would be simply $\vr^{(3)}=A\vr^{(3|1)}+B\vr^{(3|2)}$ for some matrices $A$ and $B$.

However, our illustrative goal with this example inference problem is to define how representations would be combined along the representations of $\vz_1$ and $\vz_2$, which requires a transformation, rather than simply assuming a population code directly for the higher-level variable. In this case we must describe a marginalization over the nuisance variables that distinguish $\vz_1$ and $\vz_2$ from $\vz_3$.

Marginalizing is more difficult in PPCs, because it is nonlinear in neural activity. Nonetheless, past work has shown that for Gaussian posteriors, quadratic nonlinearities with divisive normalization provide one good transformation \citep{beck2011marginalization}, where quadratic operations and divisive normalization are both biologically plausible and well-described neural operations. 

Putting these ideas together, the computation of $q(\vz_3|\vo_1,\vo_2)$ for PPCs would start with populations $\vr^{(1)}$ and $\vr^{(2)}$, each encoding a posterior $q(\vz_k|\vo_k)$ over the lowel-level variables given the corresponding evidence. A third higher-level population $\vr^{(3)}$ would then be driven by these lower ones as 
\begin{equation}
    r^{(3)}_\ell=\frac{{\vr^{(1)}} A^{\ell}\vr^{(1)}+{\vr^{(2)}} B^{\ell}\vr^{(2)}}{c^\ell+\va^\ell\vr^{(1)}+\vb^\ell\vr^{(2)}}
\end{equation}
where tensors $A_{ij}^\ell$ and $B_{ij}^\ell$ specify how each product of inputs $r^{(k)}_i$ and $r^{(k)}_j$ are weighted by neuron $r^{(3)}_\ell$, and vectors $\va_{i}^\ell$ and $\vb_{i}^\ell$ specify how these inputs affect the divisive normalization. All of these weights are specified by the representation of the input population \citep{beck2011marginalization} and the coupling strength between the $\vz$.

This describes the information flow from inputs to the representation of $q(\vz_3|\vo_1,\vo_2)$. To condition the {\it lower}-level variable on {\it both} observations, a PPC representation would repeat this process to update the population $\vr^{(1)}$, once again using quadratic operations and divisive normalization. This iterative updating defines a message-passing algorithm that converges to an equilibrium representation which reparameterizes the sufficient statistics of the joint distribution in terms of sufficient statistics for its marginals \citep{wainwright2003tree}. Since the neural activity in a PPC represents these statistics, the relevant dimensions of the neural activity also converges \citep{vasudeva2016inference}.

\subsubsection{DDC}

Unlike marginalization and chain inference, which can be easily implemented in the DDC framework, the cue combination needs elaborate computations \citep{sahani2021lecture}. Various DDC-based methods have been proposed for implementing cue integration; here, we present a proportionality relationship for the DDC of the posterior distribution which is derived through the expansion technique described in \ref{sec:DDCsIntro}. 
Let $r_i^{(1)} = \int p(z_1|o_1)\phi_i^{(1)}(z_1) d z_1$, $r_j^{(2)} = \int p(z_2|o_2)\phi_j^{(2)}(z_2) d z_2$, and $r_k^{(3)} = \int p(z_3|o_1,o_2) \phi_k^{(3)}(z_3) d z_3$ be the DDC values of the posterior distributions $p(z_1|o_1)$, $p(z_2|o_2)$, and $p(z_3|o_1,o_2)$, respectively. 
In DDC-based cue combination, ideally, we would like to compute the DDC values of the posterior $p(z_3|o_1,o_2)$, i.e. $r_k^{(3)}$, using the DDC values $r_i^{(1)}$ and $r_j^{(2)}$. 
It can be shown that $r_k^{(3)}\propto \int f_k(z_1,z_2)p(z_1|o_1)p(z_2|o_2) d z_1 d z_2$, where $f_k(z_1,z_2) = \int \frac{p(z_3)p(z_1|z_3)p(z_2|z_3)}{p(z_1)p(z_2)} \phi_k^{(3)}(z_3) d z_3$.
By approximating $f_k(z_1,z_2)$ through a bilinear expansion, $f_k(z_1,z_2) \approx \sum_{i,j} c_{i,j,k}  \phi_i^{(1)}(z_1) \phi_j^{(2)}(z_2)$, we can find the proportionality relationship $r_k^{(3)} \propto \sum_{i,j} c_{i,j,k} r_i^{(1)} r_j^{(2)}$.
Using a similar approach, we derive proportionality relationships for the marginal DDC values of $z_1$ and $z_2$. 
Let $\widehat{r}_k^{(1)} = \int p(z_1|o_1,o_2)\phi_k^{(1)}(z_1) d z_1$ denote the DDC of the marginal distribution $p(z_1|o_1,o_2)$.
We can show that $\widehat{r}_k^{(1)} \propto \sum_{i,j} \widehat{c}_{i,j,k} r_i^{(1)} r_j^{(2)}$, where $\widehat{c}_{i,j,k}$ are the expansion coefficients of $g_k(z_1,z_2) = \phi_k^{(1)}(z_1)\int \frac{p(z_3)p(z_1|z_3)p(z_2|z_3)}{p(z_1)p(z_2)} d z_3$.

\textbf{Note}: In this implementation, the bilinear expansion is a challenging task. Moreover, the DDC values of $p(z_3|o_1,o_2)$ are merely proportional to the weighted sum of the product of DDC values $\vr^{(1)}$ and $\vr^{(2)}$, and the normalization factor is not easy to calculate. To deal with these limitations, other methods have been suggested for inference and learning in hierarchical models, such as DDC-Helmholtz machine with wake-sleep algorithm \citep{vertes2018flexible} (see section \ref{sec:DDCsIntro} for more information).

\subsubsection{Neural sampling}
Simple Langevin sampling from the posterior over latents $\vz$ given observations $\vo$ takes the form of stochastic dynamics of the form:
\begin{equation}
    \dot{\vz} = -\nabla \log p(\vz|\vo) + d\epsilon
\end{equation}
where $d\epsilon$ is Brownian noise. Given the factorization of the posterior
$p(\vz|\vo) = \frac{1}{Z} p(\vo_1|\vz_1) p(\vo_2|\vz_2) p(\vz_1|z_3) p(\vz_2|z_3) P(\vz_3)$, 
with $Z$ denoting the normalizing constant, the logarithm translates into dynamics that additively combine the contribution of each element. Assuming a one-to-one map between neurons and latent variables 
\begin{eqnarray}
 \dot{z}_1  &=&-\log p(\vo_1|z_1) - \log p(z_1|z_3) \rightarrow  \dot{r}_1= f_1(\vo_1) + g_1(\vr_3)  \\
 \dot{z}_2  &=& -\log p(\vo_2|z_2) -\log p(z_2|z_3) \rightarrow  \dot{r}_2 = f_2(\vo_2) + g_2(\vr_3)\\
 \dot{z}_3 &=& - \log p(z_1|z_3) -\log p(z_2|z_3) -\log(\vr_3) \rightarrow  \dot{r}_3 = g_3(\vr_1,\vr_2,\vr_3)
\end{eqnarray}
These dynamics involve feedforward inputs $f_1(\cdot)$ to the neurons/subpopulations $\vr_1$ and $\vr_2$, respectively, with recurrent interactions  $g_i(\cdot)$ structured by the information flow described in Fig.~\ref{fig:cuecombination}b. The exact form of these functions depends on the specifics of the graphical model, e.g.\ in the linear gaussian case feedforward input effects would be linear and the recurrent interactions would implement a stochastic linear dynamical system. 
More broadly, it is important to note that within the sampling framework inference is not naturally thought of as a transformation of representations but rather as a collection of recurrently interacting dynamical systems nodes which jointly represent one posterior distribution; marginals of that posteriors available by reading out information from single nodes in this system.

\section{Interpretation of existing data}
\label{sec:interpretationOfNeuralData}

Distinguishing between different coding schemes using empirical data -- both behavioral and neurophysiological -- is complicated by the fact that the predictions generally depend not only on the coding scheme that relates a posterior probability to a neural response, but also on the generative model, $p(o,z)$, that is being used by a Bayesian brain. While it is possible to empirically evaluate such `complete models', consisting of both an assumption about the $p(o,z)$ and the coding scheme, it is currently unclear whether there are empirical signatures that can distinguish between coding schemes irrespective of the assumed generative model. That is, prediction failures associated with any given coding scheme can be attributed to incorrect choice of generative model or, equivalently, incorrect assumptions about either the identity of the latent variables represented by a given population.  In the next sections we will elucidate this fact by reviewing classic empirical observations and summarizing how they can be explained assuming different coding schemes, often involving different assumptions about the underlying generative models. 

It is important to note that these theories only put constraints on, or make predictions for, a subset of the observable biophysical properties. For example, if the activity of only a subset of neurons represent posteriors, with other neurons performing auxiliary computations (e.g.\  as in \citep{pecevski2011probabilistic,aitchison2016hamiltonian,echeveste2020cortical}), then this will pose the extra empirical challenge of identifying those neurons. Similarly, if e.g. parameters of distributions represent low-dimensional projections of high dimensional neural activity, then neural activity in directions that are orthogonal to those projections will not be constrained. In general, this caveat is a special case of the general neural coding question that asks what aspect of neural activity is computationally relevant, often applied contrasting membrane potentials with spike times or firing rates \citep{dayan2005theoretical}.

The degeneracy that arises from the possibility of the different model components of probabilistic computations to trade off against each other suggests deeper theoretical work into `equivalence classes' of different models that may all be compatible with the same biophysical system, yet involve different generative model -- neural coding pairs, or pertaining to different aspects of neural activity (also see \citep{shivkumar2018probabilistic,lange2023bayesian}).

\subsection{Tuning functions and their modulations}

\subsubsection{Tuning to a single stimulus dimension} 

When the average response of a neuron changes as a function of some variable, $s$, it is said to be `tuned' to $s$. Typically, the considered variables are experimenter-defined, such as the orientation of a visual image on the retina, or frequency of an auditory stimulus. In the context of probabilistic inference, tuning arises when the neuronal response represents the posterior over latent variables $\vz$ that depends on $s$, and that dependency changes the average response. 
In general, the tuning function is the consequence of both the coding scheme (how the response depends on the represented posterior), and how the internal variable $\vz$ depends on the experimenter chosen variable $s$.

Each of these coding schemes predicts that neurons are tuned to the represented latents, $\vz$.  If $s$ parameterizes a subspace of $\vz$ then some of the neurons representing $p(\vz)$ will also be tuned to $s$.  Moreover, if the latents are a deterministic function of $s$, i.e. $\vz=f(s)$, then not only will the population of neurons be tuned to both $\vz$ and $s$, but also the form of the probabilistic neural code (PPC/DDC/NSC) for $\vz$ will be inherited by $s$.  This suggests that complete knowledge of the relationship between represented latents $\vz$ and laboratory variables $s$ is not required for detecting the coding scheme.

\subsubsection{Scaling of tuning curves with other stimulus parameters that influence uncertainty}

\paragraph{Empirical observation: }Two of the most-studied tuning curves, those to orientation in area V1, and those to motion direction in area MT, have been shown to be `invariant' to the key stimulus aspects believed to influence the brain's uncertainty about the respective tuning variable: image contrast for orientation, and motion coherence for motion direction \citep{}. Invariant in this context means that the \emph{shape} of the tuning curve is approximately invariant, and that changes in contrast and coherence have an approximately multiplicative effect on their magnitude across the entire stimulus range. From the perspective that sensory neurons `represent' particular aspects of the input, this feature of the data appears paradoxical: while it is plausible that the neurons whose preferred stimulus is closest to the correct one would increase their firing with increasing certainty about the correct value, it is less clear why the same would be true for neurons representing stimulus values that have become less likely with increasing contrast or motion coherence \citep{}.

\paragraph{General probabilistic interpretation:} In general, the shape and scaling of tuning curves with respect to some variable $s$ will depend on the generative model defining the posterior over $z$, and the neural encoding scheme. 

\paragraph{PPC interpretation:} 
When visual contrast simply scales neural tuning over another feature such as orientation or motion direction, the PPC's logarithmic relationship between activity and posterior means that the posterior becomes narrower as the tuning amplitude rises. 
No study to date has investigated the tuning curves implied for a PPC with a generative model for e.g. natural images.

\begin{figure}[ht!]
	\centering
	\includegraphics[width =0.9\textwidth]{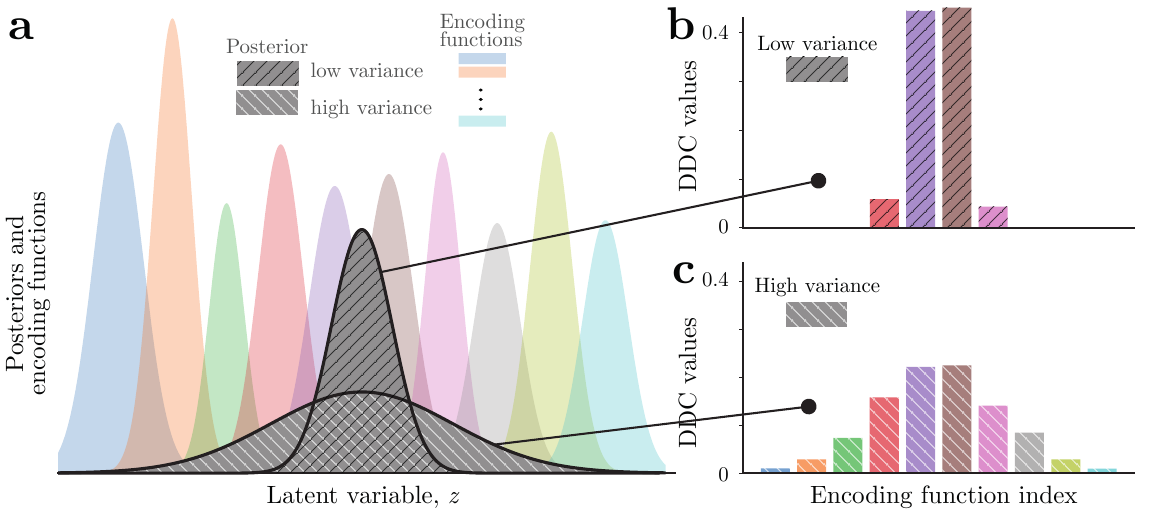}
	\caption{DDC sparsity inversely correlates with posterior variance. ({\bf a}) DDC encoding functions and two posterior distributions with low and high variance. ({\bf b}) DDC values of the posterior distribution with low variance. ({\bf c}) DDC values of the posterior distribution with high variance.}
	\label{fig:DDC_Variance}
\end{figure}

\paragraph{DDC interpretation:} 
One of the key features of the DDC representation is the modulation of population sparsity by uncertainty. When the variance (uncertainty) of the posterior distribution increases, more DDC encoding functions overlap with the distribution and the sparsity of activity reduces. In other words, the diversity of neuronal activity increases with uncertainty (Fig.\ref{fig:DDC_Variance}) \citep{ujfalussy2022sampling}. 

It is also important to highlight the distinction between the DDC encoding functions and tuning functions. To compute the tuning function of a neuron, the experimenter sweeps over the parameter of interest $\vs$ (e.g. the orientation of the grating) and measures the firing rate of the neuron. The value of the tuning function of a DDC neuron at $\vs$ is equal to the expected value of the DDC encoding function $\phi(\vz)$ with respect to the posterior distribution $p(\vz|\vo)$, where the sensory observation $\vo$ is a function of $\vs$.

The DDC framework suggests that decreasing the uncertainty of the posterior distribution should narrow the tuning functions of individual neurons. However, it has been observed that in V1, the tuning functions over the orientation are contrast-invariant, and decreasing the contrast does not broaden the tuning. Nevertheless, we should note that the generative model and the latent variables over which the DDC is defined play a critical role in this analysis; for example a multiplicative contrast term in the generative model results in a different posterior over the latent variables which correspond to the coefficients of the Gabor basis functions. 
This area warrants a more comprehensive investigation, and it is of great importance to study the impact of different generative models (e.g. for natural images) on the modulation of tuning functions in the DDC framework.

\paragraph{Sampling interpretation: } Two principal generative models for natural images have been shown to produce tuning curves to orientation (as well as other dimensions like spatial frequency) that approximately scale with contrast: Gaussian scale-mixture models under the assumption that latents are represented by membrane potentials \citep{orban2016neural}, and linear Gaussian models with binary latents represented by spikes (Chattoraj et al. COSYNE 2016) \citep{shivkumar2018probabilistic}.

\subsection{Neural variability} 
\label{sec:variability}

Both membrane potentials and spiking responses of sensory neurons are known to be variable, even to repeated presentations to the same external stimulus \citep{tolhurst1983statistical,de1997reproducibility}. Regardless of whether this variability is pure noise that is corrupting the underlying signal, or serves a computational function, it may be helpful for distinguishing between the different proposals described above.

\subsubsection{Empirical observations}

\paragraph{Neural variance}

Under natural conditions, neural responses are highly variable \citep{tolhurst1983statistical}. This variability is different in different systems: in the retina, variability in spike counts is roughly as low as possible \citep{berry1997structure}; in the early auditory system, sound direction is computed using enormous and extremely reliable synapses \citep{joris2018calyx}; but in the cortex, variance is higher, and is observed to grow roughly proportional to the mean \citep{tolhurst1983statistical,de1997reproducibility}. Indeed, the Fano factor, the ratio of the stimulus conditioned variance to the mean, is approximately one for most neurons in sensory cortical areas \citep{rieke1999spikes}. This relationship is maintained across a wide range of stimulus contrast levels (a proxy for input information in visual tasks) so that higher contrast is associated with higher firing rates and more variability.  There also appears to be an increase in variability from lower level to higher level areas \citep{kara2000low} with at least some of that increase accounted for by variability shared among neurons \citep{goris2014partitioning}.  In the absence of a sensory stimulus neural variability is higher and exhibits a marked decrease at the time of stimulus stimulus onset \citep{churchland2010stimulus} modulated by contrast in early visual areas. These findings apply both to membrane potentials and firing rate \citep{Finn2007emergence,savin2014spatio,orban2016neural,hennequin2018dynamical}.

\paragraph{Neural covariance}

Neural responses often covary across neurons. The precise pattern of covariability may be important for their information content and computational function.
Variation can be decomposed into different aspects: we expect some variation simply because the external input varies. This is often called `signal correlation'. Variability across repeated presentations of identical sensory inputs is called `noise correlations' \citep{gawne1993independent,averbeck2006effects}.  Here, the scare quotes to serve as a reminder that noise may be a misnomer as these fluctuations could be a result of useful computational mechanisms, fluctuations in attention, or context effects that were not held fixed during the experiment.  Indeed, much of the interest in neural covariation focuses on noise correlations precisely because they reflect the internal computations and structure of the neural code rather than merely the external drive to a circuit \citep{savin2014spatio,lange2017characterizing,ruff2018cognition}. 

Empirically, the noise covariance between two neurons typically grows with their mean responses, just like the variance for cortical neurons.  The covariance reflects both the relationship between the neurons and scale of each neuron's relationship. The correlation coefficient is one way to approximately isolate relationship between the neurons from the modulation of each neuron separately (although see \citep{de2007correlation, pitkow2012decorrelation}).
The pattern of correlations is often related to the neural tunings: similarly tuned neurons often exhibit greater correlations \citep{zohary1994correlated,averbeck2003neural,nardin2023structure}, although there is substantial diversity around this pattern.  Furthemore, some studies report changes in correlation with the stimulus \citep{ohiorhenuan2010sparse,ponce2013stimulus}, motor outputs \citep{dadarlat2017locomotion}, brain state \citep{ecker2014state}, and attention \citep{cohen2009attention}.

Noise covariance affects the information content that can be optimally extracted from a neuronal population \citep{zohary1994correlated,abbott1999effect,averbeck2006neural}.  When signal covariance and noise covariance are similar, as observed in cortex, the resulting neural code is highly redundant with respect to the stimulus of interest\citep{moreno2014information,ecker2014state}.  While this redundancy appears limiting, it may serve a computational purpose (e.g. \citep{lange2022task,nardin2023structure,haimerl2023targeted}).

\subsubsection{Interpretation}

There are two principal ways in which neural variability can arise among neural responses encoding a posterior belief: variability in the posterior itself, and a stochastic encoding of a fixed posterior \citep{lange2022task}. Since the latter is shared by all encoding schemes, we will discuss it first. Trial-by-trial variability in the posterior can arise even when the experimenter-controlled stimulus is kept constant as the result of external and internal sources of variability that can be mapped on to different aspects of computing a specific posterior: variability in the observation process (e.g. small eye-movements), variability in the likelihood computations, and variability in representation of prior expectations.  
Likelihoods and priors may be variable due to an approximate model of the outside world \citep{beck2012not}, approximate computations (e.g. implementations in stochastic neural circuits, or computations that only converge asymptotically). 

Under some circumstances, e.g.\ in the context of learning a behavioral task, it has been possible to characterize the nature of the neural covariability due to the variability in the posterior, and to derive empirical predictions that match that reported in existing studies \citep{lange2022task}.

\paragraph{Interpretation of neural variability by PPCs}

Unlike sampling based codes, neural variability does not play a computational role in PPCs.  However, PPCs do make strong predictions for the relationship between tuning curves and the covariance structures (Eq. \ref{decoders}) in the presence of nuisance parameters.  While there are many ways to satisfy this relationship, one oft cited way is when Fano Factors are constant (but not necessarily one) for all values of the stimulus and nuisance parameters.  As a result, the ubiquity of Fano factors near one across all stimulus conditions is often cited as evidence in favor of PPCs despite the fact that variable Fano factors can also be consistent with PPCs.  

\paragraph{Interpretation of neural variability by DDCs}

Like for PPCs, in DDC, variability does not serve a computational function but is considered noise that contaminates the signal present in the firing rates.
As noted in \ref{sec:DDCsIntro}, various noise models can be incorporated to model neural variability in the DDC framework, including additive noise (with independent or correlated components), and Poisson noise \cite{zemel1998probabilistic}.

\paragraph{Interpretation of neural variability by NSCs}

Neural variability is a necessary consequence of sampling-based inference: neural responses are variable since they are directly related to samples, which vary stochastically over time, with the amount of variability directly related to the uncertainty in the underlying beliefs.

For models based on continuous latent variables, this view predicts a dissociation of mean and variance reflecting underlying beliefs that can change in both mean and uncertainty. Importantly, how both depend on external stimulus parameters depends critically on the assumed generative model, $p(\vo|\vz)$. For instance, \citep{orban2016neural,festa2021neuronal} found that the stimulus-dependence of spiking responses and membrane potentials of V1 neurons were compatible with the assumption that membrane potentials represent samples from mixture variables in a Gaussian scale mixture model.

For models based on binary latent variables, mean and variance are more tightly coupled since the variance of a binary variable is simply $p(1-p)$ where $p$ is the probability of the variable being 1. For small $p$, and summing over many independent samples, the distribution over the count is approximately Poisson, suggesting that individual spikes (and absences of spikes) may be interpretable as samples from distributions over binary $\vz$ \citep{buesing2011neural,shivkumar2018probabilistic}. While the Fano Factor of independent binary samples is $1-p$, i.e. sub-Poisson, samples that are generated using an MCMC algorithm often have positive autocorrelations. This will increase the variability of the number of spikes counted over an extended time window, making the resulting variability  compatible with empirical observations from cortex \emph{independent} of the specific nature of $\vz$ and $p(\vo|\vz)$.

Neural sampling further predicts that whenever the posterior over different latent variables, $p(z_1,z_2|\vo)$ is correlated, then this dependency in the posterior should directly be expressed in neural co-variability between the neurons representing $z_1$ and $z_2$, respectively. Starting from this insight, several studies have derived concrete predictions for noise correlations and choice correlations in the presence and absence of behavioral tasks \citep{haefner2016perceptual,bondy2018feedback,banyai2019stimulus}. However, it is important to note that the trial-by-trial variability in the posterior (e.g. due to input noise and approximations as noted above) is closely related to the shape of the posterior itself. As a result, at least qualitatively, these predictions are shared by any coding scheme \citep{lange2022task} and, without a more quantitative analysis (see e.g. \citep{ujfalussy2022sampling}), cannot be taken as direct evidence for the neural sampling hypothesis.  Finally, the alignment of stimulus and noise correlations arises either as a consequence of approximate inference or due to a separate encoding process, e.g.\ in  distributed sampling \citep{savin2014spatio}.

\subsection{Relationship between spontaneous and evoked neural activity}

\subsubsection{Empirical observations}

Neurons in sensory cortex are active even in the absence of external inputs \citep{faisal2008noise}. This spontaneous activity does not appear random but instead appears to be structured similar to activity evoked by external sensory inputs \citep{tsodyks1999linking,kenet2003spontaneous,fiser2004small,luczak2009spontaneous}. Furthermore, the statistical structures of the spontaneous and the evoked activity appear to converge over the course of development \citep{berkes2011spontaneous}, providing a constraint on models of probabilistic computations in the brain (but see \citep{avitan2022not}).

\paragraph{Interpretation of the relationship between spontaneous and evoked activity in NSCs}

Under the assumption that complete darkness is interpreted by the brain's internal model as the absence of information about the internally represented variable, $\vz$, the likelihood, $p(\vo|\vz)$ is flat, and the posterior, $p(\vz|\vo)\propto p(\vo|vz)p(\vz)$, should equal the prior. If neural activity represents the posterior, 
spontaneous activity should therefore reflect the brain's prior. 
Furthermore, inference in a well-calibrated generative model requires that the average posterior matches the prior, $p(\vz)=\int p(\vz|\vo)p(\vo){\rm d}\vo$.
Under the assumption that neural activity represents samples from the posterior, the distribution over spontaneous activities is therefore predicted to match the distribution over activities evoked by natural stimuli, $\vo$, provided that they are presented in proportion to their natural occurrence, $p(\vo)$ -- as tested and confirmed by \citep{berkes2011spontaneous}.

\paragraph{Interpretation of the relationship between spontaneous and evoked activity in DDCs}

Since DDCs propose that neural responses represent extended moments of the posterior distribution, and therefore are linear functions of the posterior (`linear distributional codes' \citep{lange2022task}), the calibration argument described above predicts that average spontaneous activity equals average evoked activity. If moments are represented by average neural activity, i.e. firing rates, then this implies that the spontaneous firing rate should equal the average evoked firing rate. Note that this prediction is only a special case of the test in \citep{berkes2011spontaneous} who found that the \emph{distribution} over responses (spikes) equalled the \emph{distribution} over evoked responses to natural stimuli. Assuming the DDC representation has, for instance, Poisson variability around the firing rate, this will not in general be compatible with the \citep{berkes2011spontaneous} observation.

\paragraph{Interpretation of the relationship between spontaneous and evoked activity in PPCs}

For PPCs, as with DDCs and NSCs, spontaneous activity is assumed to represent the prior distribution over the relevant latent variables. However, due to PPCs' nonlinear relationship with the encoded distribution, average evoked activity is not expected to be related to spontaneous activity.

Energy efficient PPCs prefer to represent priors with low levels of neural activity to conserve spikes when there is no information to report.  This static consideration, however, is overly simplistic, as it fails to take into account the dynamics needed to implement inference. 
PPCs can straightforwardly be adapted to implement predictive coding algorithms, further increasing efficiency when performing inference on hierarchical generative models.  In this case, populations of neurons represent residual likelihoods rather than posterior distributions.  Associated patterns of activity that represent these probabilistic error signals exhibit greater transient variability in the presence of noise generated by feed forward and feedback connections.  Moreover, dynamics designed to implement probabilistic reasoning tend to turn that variability into patterns of activity that qualitatively look like patterns of activity associated with posterior distributions.  Thus, while there is not necessarily a relationship between spontaneous and evoked activity in a PPC, there are many dynamical systems utilizing PPCs that exhibit strong transient patterns of activity that resemblance evoked activity \citep{grabska2013demixing}.

\subsection{Oscillations}

Oscillations are a ubiquituous feature of cortical activity \citep{Buszaki2004neuronal}; however it is currently unknown to what they can constrain the neural implementation of probabilistice inference. On one hand they are predicted by fast algorithms implementing neural sampling using non-normal dynamics in models of hippocampus \citep{savin2014optimal} and sensory cortex\citep{aitchison2016hamiltonian,echeveste2020cortical}.
On the other hand, it has been recently shown that by modeling the hippocampal formation as a Helmholtz machine, theta oscillations can be used to mediate the wake-sleep algorithm \citep{george2024generative}. As a result, it can be suggested that the implementation of the DDC framework through the wake-sleep algorithm in a Helmholtz machine might be compatible with neural oscillations; further analysis is required to investigate this scheme.
No studies currently exist on their compatibility with PPCs. 

\subsection{Neural--behavioral correlations}
Many experiments have measured behaviors that accord with behaviors based on probabilistic inference \citep{fiser2010statistically,pouget2013probabilistic}. Of course this does not explain how the brain accomplishes this. To understand the neural basis of such behaviors, neuroscientists have examined whether neural representations of sensory uncertainty are related to actions. This is often accomplished by directly examining the relationship between neural activity and behavior.

\subsubsection{Empirical}

One common approach to measuring the relationship between neural activity and behavior is measuring the correlation between choices and individual neural responses. This correlation is known as choice probabilities \citep{britten1996relationship,haefner2013inferring} or choice correlations \citep{pitkow2015can,clery2017decision,yang2021revealing,chicharro2021stimulus}. A second approach is to predict (decode) behavior from the activity of a neural population (e.g. \citep{mante2013context} and many others). A more sophisticated variant of this approach is to decode from the neurons targeted latent properties hypothesized to be important for behavior, and then ascertain whether those decoded quantities predict behavior \citep{shahidi2019population,wu2020rational}. A recent paper used this approach to decode uncertainty about the stimulus from trial-by-trial fluctuations in neural activity, and found that this decoded uncertainty predicted shifts in decision criteria
\citep{walker2020neural}.

\subsubsection{Interpretations}

As explained in the variability section above, trial-to-trial variability in the observations will induce variability in the posterior over $\vz$ and hence behavior, inducing correlations between neural representations of $p(\vz|\vo)$ and behavior -- regardless of the neural coding scheme. Interestingly, after learning, the structure of these correlations is approximately the same regardless of neural code \citep{lange2022task}. 
As a consequence, work showing agreement between empirical correlations \citep{haefner2016perceptual,bondy2018feedback}, or decoded latents \cite{walker2020neural}, and predictions of probabilistic inference models based on particular codes, can only be taken as evidence in favor of the representation of posteriors rather than any one particular coding scheme over another.

\subsection{Behavioral data}

There are at least two ways in which behavioral data can constrain models of probabilistic computations in the brain. First, if behavior is close to optimal, then this places a constraint on the brain's internal model insofar as that the task model must be a special case of the brain's internal model. 
Second, any deviations in behavior from optimality place constraints either on the brain's internal model and/or the approximate inference algorithm. Plausible deviations of the internal model may result from being adapted to natural inputs as opposed to task-specific ones, or due to incomplete learning of the task. In those situations, Bayesian inference makes predictions about the direction of behavioral change in e.g.\ perceptual learning paradigms. Alternatively, even if the internal model is correct, the specific approximate inference algorithm employed by the brain will lead to deviations from optimal behavior. Such deviations will generally depend on the specific algorithm, and thereby observed behavior placing constraints on which algorithm is employed by the brain.

To test theories of probabilistic computations in the brain using neural data requires the specification of the link between computational quantities like samples or parameters, and neural responses (see \ref{sec:link-r-q}). In analogy, testing the same theories using behavioral data requires a link between posteriors and actions. While much work exists on the nature of this link (e.g.\ \citep{kording2007decision}), we consider this beyond the scope of this manuscript, and will only briefly describe selected attempts in the hope to encourage further research in that direction and describe caveats.

\subsubsection{Behavioral bias}

As an example, \citeauthor{haefner2016perceptual} found that hierarchical inference by sampling in the context of a sequential evidence integration task led to an overweighting of evidence presented early in a trial\citep{haefner2016perceptual} . A follow-up study elaborating on this initial finding discovered that rather than being specific to sampling, this bias was also predicted by a variational inference model based on a parametric representation \citep{lange2021confirmation}. By interpolating between two related but different tasks, and by explaining data across both tasks with the same generative model structure, and the same inference algorithm, this study uses generalization across tasks as a way to address a general critique of probabilistic approaches: it is possible to explain any behavior as optimal inference on some generative model. However, this can be seen as a form of overfitting to a task, yielding a generative model that may not generalize to other tasks.

\subsubsection{Behavioral variability}

Just as for neural responses, variable behavior may arise as the result of variability in the posterior due to uncontrolled variability in the observations, or variability in the neural encoding or computation of the posterior \citep{drugowitsch2016computational,lange2021confirmation,shivkumar2022inferring}. Furthermore, the variability in the posterior itself may be magnified by a mismatch between the internal model and the model generating the observations \citep{beck2012not}. 

Much like neural variability, human and nonhuman primate perception has been shown to be variable, even for constant external inputs. Examples are images with ambiguous interpretation (e.g. the vase/face image), and dichoptic stimuli that induce perceptual switching between the image shown to the left and the right eye \citep{blake2002visual}.  \citep{gershman2012multistability} showed that sampling from a probabilistic model of bistable inputs implied a distribution over dominance times that qualitatively matched empirically observed distributions. 
\citep{moreno2011bayesian} showed that under the assumption that the posterior is represented as a linear PPC, an attractor network could generate samples from the posteriors that matched empirically observed distributions.

\subsection{Does the entirety of the considered empirical data favor one of the neural codes?}

Overall, most of the observations considered above can qualitatively be explained by all of the proposals. The only exception may the observation that spontaneous activity is very similar to the average evoked activity: while this is directly predicted by neural sampling, it seems significantly harder to explain by DDCs and PPCs.

\section{Recommendations for future studies}
\label{sec:future}

\paragraph{Testing for relationships between the neural representations for different posteriors:} While systems neuroscience has traditionally focused on stimulus-response relationships, focusing on the relationships between the responses holds the promise of empirical tests of the different coding schemes that do not require explicit knowledge of the $\vz$ that is being represented. A key example of this approach exploits the calibration property of probabilistic models: that the average posterior should equal the prior. This was tested by \citep{berkes2011spontaneous} who compared spontaneous activity in area V1 to average evoked activity and found that they are statistically indistinguishable in mature ferrets. This result is predicted for any system that represents posterior beliefs under the assumption that placing the ferrets in a completely dark room is interpreted as `no input' by the visual system. Recent work \citep{lengyel2023general} has shown how to exploit a linearity property obeyed by some neural codes (DDCs and NSCs), but not all (e.g. PPCs), to test whether neural responses to different stimuli are linearly related to each other in a way suggested by the underlying posteriors to those stimuli. Such an approach may allow for the development of a method to compare brains and probabilistic models that is akin to representational similarity analysis (RSA \citep{kriegeskorte2008representational}) that allows for the empirical test of generative models and neural codes requiring weaker assumptions than currently needed by explicating a `complete' model.

\paragraph{Generalization across multiple experiments and datasets: }Given many possible choices for latent variables $\vz$, the generative model linking them to experimental variables $p(\vz,\vo),\vs$, and the concrete mapping between the posterior and the neural responses $\vr$, no single experimental dataset will be enough to constrain all of these degrees of freedom. However, we can make progress by focusing on the generalization properties of the probabilistic model. For data fitting to go beyond `Bayesian just so' storytelling \citep{bowers2012bayesian}, the same set of model choices should be able to explain \emph{all} aspects of the data, not just a subset (e.g. tuning functions, their changes with uncertainty, or a given feature of response covariability, structure of spontaneous response covariability, behavioral response changes with uncertainty, etc). Testing for consistency of data-constrained probabilistic quantities across tasks (prior, latents, mapping to neural activity) thus seems to be a productive approach to validate such hypotheses, akin to behavioral-level attempts at constraining probabilistic descriptions of perception at the behavioral level \citep{maloney2002statistical,houlsby2013cognitive}.

\paragraph{Develop a quantitative benchmark for the comparison of probabilistic models: } Another promising direction is to design a benchmark consisting of all relevant aspects of neural activity in one cortical area (e.g. membrane potentials, spike times, and spike rates), for a specific set of stimuli and behavioral contexts. This would facilitate fair comparisons when comparing different models and encoding schemes, and may accelerate progress for the same reasons benchmarks have been helpful in machine learning. Recent work has taken a step in that direction by directly fitting a flexibly parameterized generative model to neural responses to natural images under the assumption of a neural sampling code \citep{shrinivasan2024taking}.

\paragraph{Causal manipulations: }
The strongest way to test our understanding of the computations performed by a neural circuit is to causally manipulate that circuit. This allows us to directly attribute computational or behavioral consequences to the manipulated property. Modern neuroscience methods afford multiple ways of performing such causal interventions, including electrical, pharmacological, or optogenetic manipulations. As the controllable spatial and temporal resolution of these manipulations increases, our interventions can be more targeted to specific activity patterns of interest. However, even coarse interventions may provide illuminating tests for discriminating between probabilistic coding schemes.

For instance, cooling or optically inactivating an area implies that the marginal belief about the corresponding variable $z_1$ is either completely confident that the latent variable is zero,  $p(z_1)=\delta(z_1)$ (for NSCs), or uninformative with $p(z_1)=\rm const$ (for PPCs). As a result, NSCs predict that the neural variability in other areas representing $z_2$ should be decreased (NSCs), while PPCs predict that the mean activity should be reduced since $p(z_2)$ will be less certain.

\paragraph{Interactions across brain areas: }
We have emphasized the importance of computations because representations do not stand on their own. Their value lies in their use. Decades of past evidence has demonstrated that brain areas exhibit both some specialization and some hierarchical structure. Consequently, we expect that representations of different latent variables in different brain areas will influence each other in predictable ways. Some of this may be discernible through observing existing correlations \citep{semedo2019cortical}. However, causal manipulations including inactivation \citep{lakshminarasimhan2018inferring}, noise injection, or patterned perturbation \citep{chettih2019single,adesnik2021probing} are valuable in distinguishing direct interactions from indirect ones or from common causes \citep{lakshminarasimhan2018inferring,das2020systematic}. For all codes, only a subspace of neural activity may dominate the encoding of the parameters of the encoded probability distribution, so only perturbations that affect those dimensions should affect computations in other brain areas. In PPCs and DDCs, these are dimensions that project onto the sufficient statistics of the posterior. In sampling, these are dimensions that contribute to fluctuations in latent variables. For example, if multiple neurons represent the same variable, such that their population mean represents the sample, then population dimensions that increase some neurons' firing while decreasing others' will have no effect. Experiments that measure and then manipulate (or track noise in) these dimensions will allow the testing of whether these dimensions affect downstream computation in the manner predicted by each theory.

\paragraph{Theoretical work on similarities and differences between the coding schemes:}
Theoretical work may find that these models are formally equivalent under some conditions. For example, recall that all of these models make predictions that depend critically on committing to assumptions of a generative model, including what latent variables $\vz$ the distribution is over, what properties of neural activity $\vr$ encode them, and  how these are connected to observations $\vo$. For example, \citep{lange2023bayesian} showed that sampling over basis function amplitudes can manifest as a PPC over orientations on a coarser timescale. Different choices of experimental variables $\vs$ to probe possible latent variables $\vz$ may also lead to indistinguishable conclusions about the brain's confidence when these two types of variables are highly informative about each other. Future studies may find that such relations hold more generally, underscoring the importance of defining and comparing the key assumptions from which predictions are derived.
Deeper insights on the similarities and equivalences of different coding schemes will also yield a better understanding of the predictions on which these schemes actually disagree, and the kind of data and experiments that may be able to definitely distinguish between them.

\paragraph{Theoretical work on general framework which contains specific coding schemes as special cases: }
Following existing work, this paper treats different coding schemes as alternative hypothesis. However, it might be more productive to conceived them as special cases of a more general coding scheme. For instance, \citeauthor{lange2022interpolating} described a space in which variational and sampling-based inference represent two extreme points along a continuum of inference algorithms some of which may be a closer description of the brain's encoding scheme than either of the extremes.

\subsection{Conclusion}

There is growing empirical evidence that Bayesian inference, and Bayesian decision theory, is a useful framework for understanding human behavior. On the basis of this it is tempting to view neural activity through the same lens. However, the jury is still out whether this is a fruitful approach, and whether the Bayesian framework has predictive power for neural activity. For instance, outputs consistent with probabilistic computations emerge generically when a sufficiently flexible computational system is trained in a world where probabilistic inference is the best way to solve problems \citep{ramsey1926truth,orhan2017efficient}. However, whether the brain's implementation can be mapped onto Bayesian concepts like priors, likelihoods, posteriors, loss functions, variational parameters, moments, or samples, is less obvious. But if it can, and if these mappings generalize across sensory inputs and behavioral tasks -- an empirical question -- then this would greatly advance our understanding of the brain linking the computational level with the implementation level via the algorithmic level.

\section*{Acknowledgements}

We would like to thank the participants of the CCN/GAC Kick-off workshop, in particular the presenters Alex Pouget, Rajesh Rao, Maneesh Sahani, Eero Simoncelli, Eszter Vertes, Balazs Ujfalussy, Emin Orhan, and Richard Lange. We also thank Richard Lange, Gerg\H{o} Orban and Suhas Shrinivasan for comments on the manuscript, and M\`at\`e Lengyel for helpful discussions. 
This work was financially supported by NIH U19NS118246 (RMH,XP), NSF/CAREER IIS-2143440 (RMH), Gatsby Charitable Foundation, Simons Foundation (SCGB 543039), and NSF Award No. 1922658 (CS).

\bibliographystyle{plainnat} 
\bibliography{references.bib}

\end{document}